\documentclass[12pt]{article}
\usepackage{eurosym}
\usepackage{graphicx}
\usepackage[utf8]{inputenc}
\usepackage[T1]{fontenc}
\usepackage{indentfirst}
\usepackage[margin=0.6in,nomarginpar]{geometry}
\usepackage[final]{hyperref}
\usepackage{amsmath}
\usepackage{hyperref}
\usepackage{cite}
\usepackage{subcaption}
\usepackage{caption}
\usepackage{amssymb}
\usepackage{multirow}
\usepackage[table]{xcolor}
\usepackage{orcidlink}
\usepackage{color}
\hypersetup{
colorlinks=true,
linkcolor=blue,
citecolor=blue,
filecolor=magenta,
urlcolor=blue
}

\begin{document}


\title{Schwarzschild Black Hole Surrounded by Perfect Fluid Dark Matter in the presence of Quintessence Matter Field}

\author{B. Hamil 
 \orcidlink{0000-0002-7043-6104} \thanks{%
hamilbilel@gmail.com} , \\
Laboratoire de Physique Math\'{e}matique et Subatomique,\\
Facult\'{e} des Sciences Exactes, Universit\'{e} Constantine 1, Constantine,
Algeria. \and B. C. L\"{u}tf\"{u}o\u{g}lu 
 \orcidlink{Orcid ID : 0000-0001-6467-5005} \thanks{%
bekir.lutfuoglu@uhk.cz 
 (Corresponding author)} , \\
Department of Physics, Faculty of Science, University of Hradec Kralove, \\
Rokitanskeho 62/26, Hradec Kralove, 500 03, Czech Republic. }
\date{\today }
\maketitle

\begin{abstract}
{In this paper, we present an exact solution for a spherically symmetric Schwarzschild black hole surrounded by perfect fluid dark matter (PFDM) in the presence of a quintessence field. We investigate the impact of dark matter on the black hole's thermodynamic and optical properties, as well as its quasinormal modes. Our analysis reveals a critical radius where the heat capacity becomes positive, indicating the thermodynamic stability of the black hole. Notably, this critical radius increases with the dark matter parameter $\alpha$. Additionally, we find that as the effects of dark matter increase, the black hole's shadow radius decreases. Using the WKB approximation, we show that the quasinormal mode spectrum differs from that of a standard Schwarzschild black hole due to the influence of PFDM. Moreover, we demonstrate that as $\alpha$ increases, both the real part and the magnitude of the imaginary part of the quasinormal mode frequencies increase. This suggests that field perturbations decay more rapidly in the presence of PFDM compared to those in a {Schwarzschild black hole}.}
\end{abstract}


\section{Introduction}

The universe is a great, mysterious puzzle. Only about 5 $\%$ of our universe consists of directly observable ordinary matter, while the rest is comprised of dark matter and dark energy \cite{Bertone2018, Huterer2018}. Although these dark sector components have not been directly observed, their existence has been inferred from various indirect findings \cite{Rahaman2010, Potapov2016}. For example, the presence of dark matter has been predicted by several phenomena including the rotation curves of spiral galaxies \cite{Rubin1970, Roberts1973, Rubin1980}, the dynamical stability of disk galaxies \cite{Einasto1974, Ostriker1973}, the large-scale structure formation of the universe \cite{Davis1985}, gravitational lensing observations by galaxies and galaxy clusters \cite{Clowe2006, Mandel2013}, the cosmic microwave background \cite{Hinshaw2013, Ade2014}, and baryon acoustic oscillations \cite{Komatsu2011}. { Numerous} models have been developed to explain dark matter itself \cite{Martino2020, Salucci2020, Filippi2020, Arbey2021, Oks2021a, Oks2023, Navarro1996, Boehm2004, Bertone2005, Feng2009, Graham2015, Schumann2019, Boyarsky2019, Qiao2021,Deur2019, Yaholom2020, Oks2020, Oks2024}. Among these, the perfect fluid dark matter (PFDM) model, which treats dark matter as a perfect fluid, presents intriguing theoretical aspects. For instance, Kiselev derived exact static spherically symmetric black hole solutions from Einstein's equations using the PFDM model \cite{Kiselev2022}. Subsequently, Li et al. utilized the logarithmic term from Kiselev’s PFDM model to analyze the asymptotic rotation curves of dark matter in halo-dominated regions \cite{Li2012}. Xu et al. generalized Kiselev's solution to Kerr-like black holes using the Newman-Janis method and explored Kerr-(A)dS black hole properties \cite{Xu2018}. Further studies examined the impact of PFDM on deflection angles, black hole shadows, naked singularities, and thermal properties \cite{Hou2018, Haroon2019, Rizwan2019, Hendi2020}. Recent research has also explored PFDM in the context of other black hole types, including Reissner-Nordström, Reissner-Nordström-AdS, Schwarzschild, Schwarzschild-dS, and non-linear magnetic-charged AdS black holes \cite{Xu2019, Atamurotov2022, Rayimbaev2021, Atamurotov2021, Ndongmo2023, Rakhimova2023, Qiao2023, Heydarifard2023, Das2024, Sapher24}.

Reliable astronomical observations put forward that our universe is expanding at an accelerating rate \cite{Riess1, Riess2, Perlmutter1999}, and dark energy, the other component of the dark sector, is thought to be behind this fact. The cosmological constant, whose observational and theoretical values differ by $120$ orders of magnitude \cite{Carroll2001}, cannot explain this expansion. Theoretical physicists proposed new models using dynamical scalar fields to describe dark energy, taking into account that the cosmological constant does not alter with time \cite{carroll1998, Khoury2004, Picon2000, Hellerman2001, Padman2002, Caldwell2002, Gasperini2002, Copeland2006}. Among different models, the quintessence matter model stands in the form where the equation of state is limited by the linear relationship between pressure, energy density and a state parameter \cite{Hellerman2001}. Using the quintessence matter field, Kiselev solved Einstein's equation and obtained a general solution form of the static spherically symmetric black hole \cite{Kiselev2003}. Subsequently, in the presence of quintessence matter new solutions of other black holes were obtained, and the impacts of the quintessence field on their thermal quantities, shadows, and quasinormal modes were discussed widely in \cite{Chen2005, Zhang2006, Chen2008, Wei2011, Thomas2012, Fernando2013, Ghosh2016, Ma2017, XuKerr, Ghaderi2018, Fontana2018, Wu2018, Liu2019, Shahjalal2019, Haldar2020, Nozari2020, Nozari2022, BB1, Ndongmo, Chen, BB2, Singh2022, Zhang, BB3, BB4, Wang2023, Hui2024, BB5, Bamba}.

Although black hole solutions have been sought for more than a century, the problem of their detection remained an open debate. Until the successful observational results of the Event Horizon Telescope were published \cite{M871, M871a, M871b, SagA1, SagA1a}, some authors suggested that not the black holes but their shadows could be detected \cite{Synge66, Bardeenk, Luminet}. According to this proposal, if a black hole is located between a light source and an observer, it should be expected that some of the light rays would be deflected by the intense gravitational field of the black hole and reach the observer. However, some photons would fall into the black hole instead of this deflection and create a dark region called a shadow. The apparent shape of the black hole is assumed to correspond to the boundary of the shadow. { By studying black hole shadows, we can probe the effects of both dark matter and dark energy on the geometry of space-time, making them important tools for investigating alternative gravity theories and understanding the universe's large-scale structure. With this motivation nowadays many papers} examine the impact the space-time geometry, dark energy, and dark matter on shadow images in the literature \cite{Tsu3, Konoplyae, WeiMann, Babar, Kumar, Singh, Panah20,  Das20, Das21, Das22, Nodehi2020, LiGuo, Zhang21, Thomas1, AtamuratovJamil2021, Rayimbaev2022, Das, Pantig2022, GuoWD, PG1, ZG1, Sunny1, Sunny2, Sunny3, Sunny4, Sunny5, Sunny6, Vir, Vir1, Biz2023, Biz2023b, Du, Akhil, Kara, Ali1, Chowdhuri, Molla, Olmo, AtamuratovJamil2023, Panah23, Panah24, Ghosh2023, Hoshimov2024, Pantig1, Pantig2, Pantig3}.


{ The detection of gravitational waves  \cite{BF1, BF2} has provided significant insights into fundamental astrophysical phenomena \cite{Maggiore2005}. Advanced observations of these waves have become essential tools for testing various theories that extend beyond general relativity, including both modified and alternative theories of gravity. Modified theories, such as $f(R)$ gravity \cite{fRrev, Nojiri2011}, Scalar-Tensor theory \cite{ScalarTensorGrav, Capozziello2011}, Einstein–Gauss-Bonnet theory \cite{EGBrev}, massive gravity theory \cite{MassGrav} etc. stem from adjustments to the Einstein-Hilbert action while maintaining a geometrical structure consistent with general relativity. In contrast, alternative theories, including teleparallel gravity \cite{TPG1, Capozziello2019, Bahamonde, TPGrev}, loop quantum gravity \cite{LQGrev}, string theory \cite{Polchinski} introduce fundamentally different geometrical frameworks and offer unique predictions for gravitational waves, particularly under extreme conditions near black holes.  Of particular interest, black holes in their final stages are theorized to emit gravitational waves as quasinormal radiation \cite{Press1971}. Mathematically, quasinormal modes are represented as complex numbers, with the real part corresponding to the oscillation frequency and the imaginary part representing the inverse damping time \cite{Mod2}. Their independence from initial perturbations allows them to be considered distinctive signatures of black holes \cite{Konoplya2011}. A substantial body of literature has utilized both analytical and numerical methods to explore the quasinormal modes of black holes in various scenarios, thereby elucidating their implications for gravitational theories and advancing our understanding of black hole dynamics within the context of contemporary astrophysical observations \cite{Zhidenko, Rincon2018, Chirenti2018, Tattersall, Myung2019, Kon2019, Salcedo2019, Kon2020, Liu2020, Jusufi2020, Roman2020, HendiJamil2020, HendiJamil2021, Kanzi2021, Anacleto2021, Gulnihal2021, Jafarzade, Ghosh2021, Okyay2022, Ovgun2022, Chen2022, Sakalli2022, Roman2023, Nesrin2023, Anacleto2023, Jha2023, Lambiase2023, Demir2023, Ghosh23, Badawi2024}.}


{ Furthermore, the thermodynamics of black holes has become a fundamental area of study, particularly following the establishment of the connection between a black hole's event horizon and its entropy \cite{Bekenstein1973, Hawking1974}. The presence of PFDM and quintessence fields modifies the energy-momentum tensor, altering the geometry of spacetime around black holes and impacting their thermodynamic properties, such as temperature, heat capacity,  and introduce new critical phenomena. These modifications could lead to generalizations of classical thermodynamic results, contributing to our understanding of the information paradox and potentially offering new perspectives on holography and quantum gravity .} 

In conclusion, studying the quasinormal modes, shadows, and thermodynamics of black holes is crucial for investigating the influence of dark matter and dark energy, testing alternative gravity theories, and deepening our understanding of black holes' fundamental properties and behavior under diverse cosmic conditions. Motivated by these {consideration}, this contribution aims to explore the thermodynamics, shadows, and quasinormal modes of a Schwarzschild black hole surrounded by PFDM, with the presence of the quintessence matter field. { To achieve these goals, the paper is organized as follows: In Sec. \ref{sec2}, the mathematical framework of the scenario under consideration is briefly formulated. In Sec. \ref{sec3}, the thermodynamics of the system is examined. The anticipated shadow images and quasinormal modes of the black hole are then presented in Secs. \ref{sec4} and \ref{sec5}, respectively, using two distinct approaches. Finally, the manuscript is concluded in Sec. \ref{sec6}. Note that throughout the manuscript, natural units are used, with
$\hbar= k_{B}=1$}.



\section{Black Hole Solution} \label{sec2}
{\ In the context of quintessence, the action associated with Einstein
gravity in the presence of PFDM can be written as \cite{Li2012, Xu2018,
Xu2019, Perez}: 
\begin{equation}
S=\int d^{4}x\sqrt{-g}\left[ {\frac{c^4}{16\pi G} R}- \mathcal{L}%
^{(PFDM)}+\mathcal{L}^{\left( {quint}\right) }\right] ,  \label{e1q}
\end{equation}%
where $G$ is Newton's constant, $c$ is the speed of light, $R$ is the scalar
curvature, $\mathcal{L}^{(PFDM)}${\ is the PFDM Lagrangian density, and $%
\mathcal{L}^{(quint)}$ is the Lagrangian density of the quintessence fields,
which is expressed in terms of the quintessential scalar field $\phi $ and
the potential $V\left( \phi \right) $ as follows \cite{Mehdi}: 
\begin{equation}
\mathcal{L}^{\left( {quint}\right) }=-\frac{1}{2}\left( \nabla \phi \right)
^{2}-V\left( \phi \right) .
\end{equation}%
Applying the variational principle to Eq. \eqref{e1q} with respect {%
to the contravariant metric tensor} $g^{\mu \nu }$, we get 
{\small 
\begin{equation}
\frac{1}{\sqrt{-g}}\frac{\delta S}{\delta g^{\mu \nu }}=\frac{1}{\sqrt{-g}}%
\left[ \frac{c^{4}}{4\pi G}\int d^{4}x\frac{\delta \left( \sqrt{-g}R\right) 
}{\delta g^{\mu \nu }}- \int d^{4}x\frac{\delta \left( \sqrt{-g}\mathcal{%
L}^{(PFDM)}\right) }{\delta g^{\mu \nu }}+\int d^{4}x\frac{\delta \left( 
\sqrt{-g}\mathcal{L}^{\left( {quint}\right) }\right) }{\delta g^{\mu \nu }}%
\right] =0,  \label{e9q}
\end{equation}%
}{\normalsize which leads to the field equation 
\begin{equation}
R_{\mu \nu }-\frac{1}{2}g_{\mu \nu }R=\frac{8\pi G}{c^4}\Big(T_{\mu \nu }^{(PFDM)}-T_{\mu \nu
}^{\left( {quint}\right) }\Big),  \label{ticom}
\end{equation}%
where the energy-momentum tensors are 
\begin{eqnarray}
T_{\mu \nu }^{(PFDM)} &=&\frac{2\delta \left( \sqrt{-g}\mathcal{L}%
^{(PFDM)}\right) }{\delta g^{\mu \nu }}, \\
T_{\mu \nu }^{\left( {quint}\right) } &=&\frac{2\delta \left( \sqrt{-g}%
\mathcal{L}^{\left( {quint}\right) }\right) }{\delta g^{\mu \nu }},
\end{eqnarray}%
in units of $\frac{4\pi G}{c^{4}}=1$. Since we consider dark matter as a type of perfect fluid, { we can represent it with an effective energy-momentum tensor. For a perfect fluid, the energy-momentum tensor is: 
\begin{equation}
T^{\mu \nu (PFDM)}=\left( \rho +p\right) u^{\mu }u^{\nu }+pg^{\mu \nu },
\end{equation}%
where $\rho $ is the energy density, $p$ is the pressure, and $u^{\mu }$ is the four-velocity. In addition, for the simplest case, we assume \cite{Li2012}, 
\begin{equation}
T_{\text{ \ }\theta }^{\theta (PFDM)}=T_{\text{ \ }%
\varphi }^{\varphi (PFDM)}=\left( 1-\delta \right) T_{\text{ \ }t}^{t(PFDM)},
\end{equation}%
where $\delta $ is a constant. To find a solution that satisfies Eq. %
\eqref{ticom}, we assume a spherically symmetric line element \cite{Xu2019,Ragil}:
{{{\normalsize \ 
\begin{equation}
ds^{2}=-f\left( r\right) dt^{2}+\frac{1}{f\left( r\right) }%
dr^{2}+r^{2}\left( d\theta ^{2}+\sin ^{2}\theta d\varphi ^{2}\right) ,
\label{30met}
\end{equation}%
or equivalently, 
\begin{equation}
ds^{2}=-\left( 1-m\left( r\right) \right) dt^{2}+\frac{1}{1-m\left( r\right) 
}dr^{2}+r^{2}\left( d\theta ^{2}+\sin ^{2}\theta d\varphi ^{2}\right) .
\end{equation}%
Replacing the metric ansatz in Eq. (\ref{ticom}), the different components of
Einstein }}}equations take the form{{\normalsize \ }}
{{{\normalsize 
\begin{equation}
G_{0}^{0}=-\frac{m\left( r\right) }{r^{2}}-\frac{1}{r}\frac{dm\left(
r\right) }{dr}+2T_{0}^{0\left( {quint}\right) }=2T_{0}^{0\left( {PFDM}%
\right) },  \label{ticom2}
\end{equation}%
}}}%
\begin{equation}
G_{2}^{2}=-\frac{1}{r}\frac{dm\left( r\right) }{dr}-\frac{1}{2}\frac{%
d^{2}m\left( r\right) }{dr^{2}}+2T_{2}^{2\left( {quint}\right)
}=2T_{2}^{2\left( {PFDM}\right) },  \label{tico2}
\end{equation}%
{{{\normalsize where 
\begin{equation}
T_{\mu }^{\nu \left( {PFDM}\right) }=diag\left( -\rho ,p_{r},p,p\right) ,
\label{eqqui}
\end{equation}%
}}}
and%
\begin{equation}
T_{0}^{0\left( {quint}\right) }=T_{r}^{r\left( {quint}\right) }=-\frac{%
3\sigma \omega _{q}}{2r^{3\left( \omega _{q}+1\right) }};T_{2}^{2\left( {%
quint}\right) }=-\frac{\left( 3\omega _{q}+1\right) }{2}T_{0}^{0\left( {quint%
}\right) }.
\end{equation}
Furthermore, in the simplest case, we consider the equation of state for the PFDM as
\begin{equation}
p=\left( \delta -1\right) \rho ,
\end{equation}%
and taking the ratio of the above Eqs. (\ref{ticom2}) and (\ref{tico2}), we
get%
\begin{equation}
\frac{1}{2}\frac{d^{2}m\left( r\right) }{dr^{2}}+\frac{\delta }{r}\frac{%
dm\left( r\right) }{dr}+\left( \delta -1\right) \frac{m\left( r\right) }{%
r^{2}}-\frac{\left( 3\omega _{q}+1\right) }{2}\frac{3\sigma \omega _{q}}{%
r^{3\left( \omega _{q}+1\right) }}+\left( \delta -1\right) \frac{3\sigma
\omega _{q}}{r^{3\left( \omega _{q}+1\right) }}=0.
\end{equation}%
The solution of this equation for $\delta =3/2$ is
\begin{equation}
m\left( r\right) =\frac{r_{s}}{r}-\frac{\alpha }{r}\ln \frac{r}{\left\vert
\alpha \right\vert }+\frac{\sigma }{r^{1+3\omega _{q}}},
\end{equation}%
where $r_{s}$ and $\alpha $ are the integration constants. To find $r_{s}$,
we consider the case where both $\alpha =\sigma =0$. Under these conditions,
we find that $r_{s}=2M$. This then determines the form of the lapse function,}
\begin{equation}
f\left( r\right) =1-\frac{2M}{r}+\frac{\alpha }{r}\ln \frac{r}{\left\vert
\alpha \right\vert }-\frac{\sigma }{r^{1+3\omega _{q}}}.\label{f}
\end{equation}

We notice that the quintessence state parameter, which varies between $-1/3<\omega_q<-1$, has a crucial impact on the {metric} function. Therefore we have to discuss black hole properties with specific values of it. To be consistent throughout the rest of the manuscript, we choose the following four values: $\omega_q=-0.35$, $\omega_q=-0.55$, $\omega_q=-0.75$, and $\omega_q=-0.95$. At first in Fig. \ref{figlaps} we depict the {metric} function with these parameters versus radius. 
\begin{figure}[htb!]
\begin{minipage}[t]{0.5\textwidth}
        \centering
        \includegraphics[width=\textwidth]{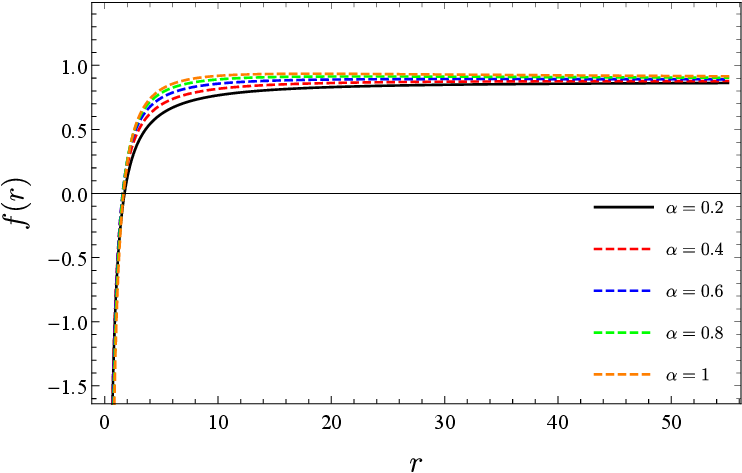}
       \subcaption{ $ \omega_{q}=-0.35$.}\label{fig:Ma}
   \end{minipage}%
\begin{minipage}[t]{0.5\textwidth}
        \centering
        \includegraphics[width=\textwidth]{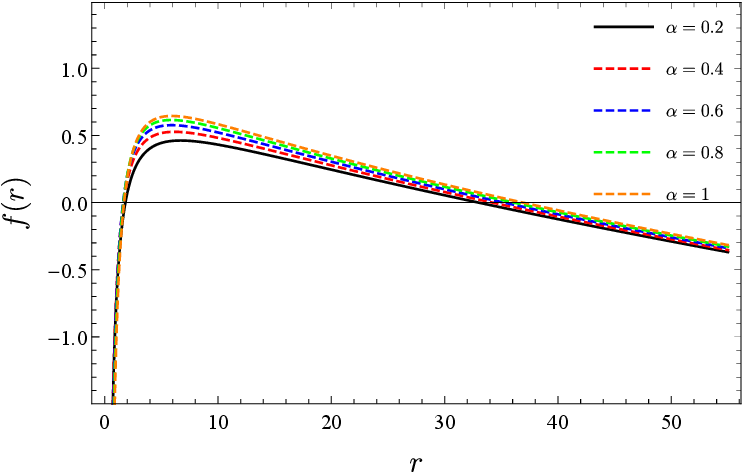}
         \subcaption{ $ \omega_{q}=-0.55$.}\label{fig:Mb}
   \end{minipage}\ 
\begin{minipage}[t]{0.5\textwidth}
        \centering
        \includegraphics[width=\textwidth]{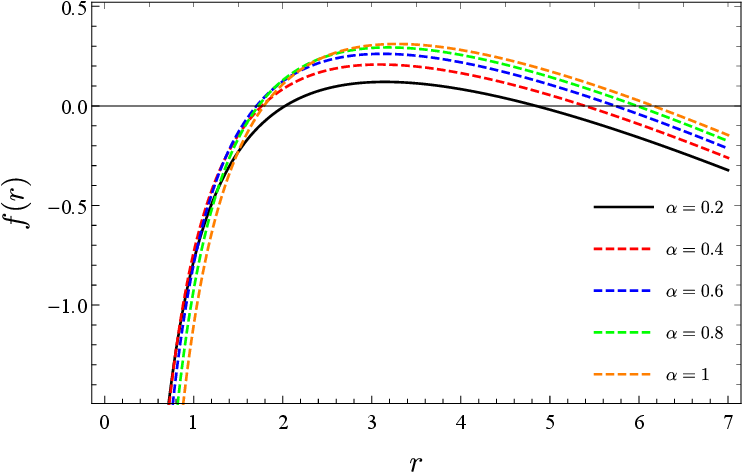}
       \subcaption{ $\omega_{q}=-0.75$.}\label{fig:Mc}
   \end{minipage}%
\begin{minipage}[t]{0.5\textwidth}
        \centering
        \includegraphics[width=\textwidth]{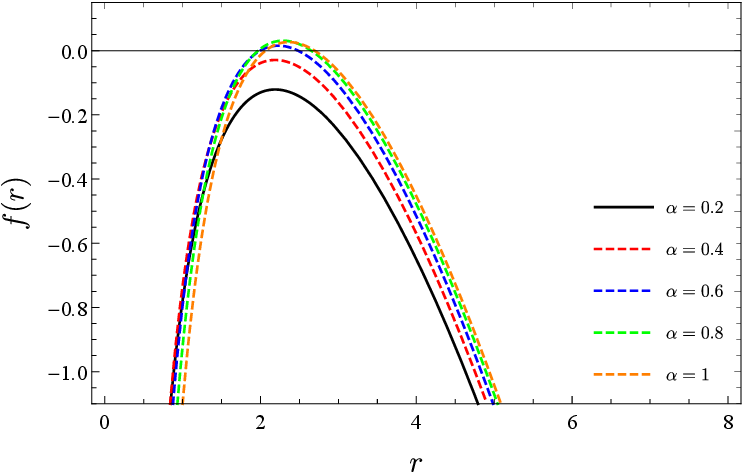}
         \subcaption{ $ \omega_{q}=-0.95$.}\label{fig:Md}
   \end{minipage}
\caption{The impact of the dark matter and dark energy on the {metric} function of Schwarzschild black hole for $\sigma=0.1 $ and $M=1$.}
\label{figlaps}
\end{figure}

\newpage
We observe that the quintessence state parameter alters the concavity of the {metric} function which is important in the context of the event horizon and naked singularity.  For example, for $\omega_q=-0.35$ there is only one horizon. In this case, the dark matter parameter does not significantly influence the horizon radius. However, for smaller quintessence state parameters the concavity increases. For instance, for $\omega_q=-0.55$, we observe { that as PFDM influence increases, the metric function $f(r)$ peaks higher and decreases more slowly with distance. All curves indicate two horizons where $f(r)=0$. Stronger PFDM effects (higher $\alpha$) appear to expand the region between these horizons and extend the black hole's gravitational influence farther into space. The graph suggests dark matter significantly alters the black hole's spacetime structure, potentially affecting its observable properties}.
For much smaller quintessence matter parameters, we see that the impact of PFDM more apparently. For example, for $\omega_q=-0.75$, the greater horizon radii occur relatively at smaller horizon values because of the larger concavity. All these cases are free of a naked singularity. However, for much smaller quintessence state parameters, for instance, for $\omega_q=-0.95$, the concavity increases so much that the {metric} function may not reach zero for any radius. In this case, naked singularity can appear. On the other hand, with greater dark matter parameters an event horizon can occur and the naked singularity problem can be resolved. It is worth noting that these generic scenarios are plotted for a random mass value, $M=1$. For a different mass value, it is possible that there is no naked singularity. 


Next, we use the horizon equation
\begin{equation}
f\left( r\right) \Big\vert _{r=r_{H}} = 0,  \label{HOR}
\end{equation}
for deriving the mass function in terms of the event horizon. We find
\begin{equation}
M=\frac{r_{H}}{2}-\frac{\sigma }{2r_{H}^{3\omega _{q}}}+\frac{\alpha }{2}\ln 
\frac{r_{H}}{\left\vert \alpha \right\vert }.  \label{m11}
\end{equation}
In Figure \ref{figmass} we demonstrate how the mass function varies with the event horizon using the parameters employed above. 
\begin{figure}[htb!]
\begin{minipage}[t]{0.5\textwidth}
        \centering
        \includegraphics[width=\textwidth]{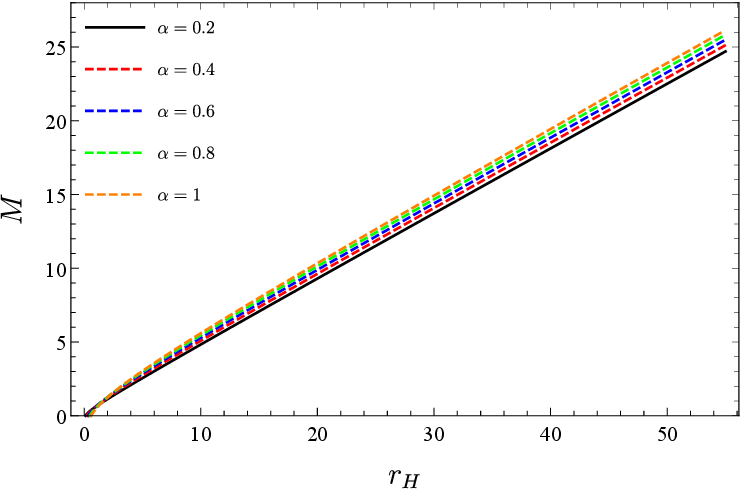}
       \subcaption{ $ \omega_{q}=-0.35$.}\label{fig:Maa}
   \end{minipage}%
\begin{minipage}[t]{0.5\textwidth}
        \centering
        \includegraphics[width=\textwidth]{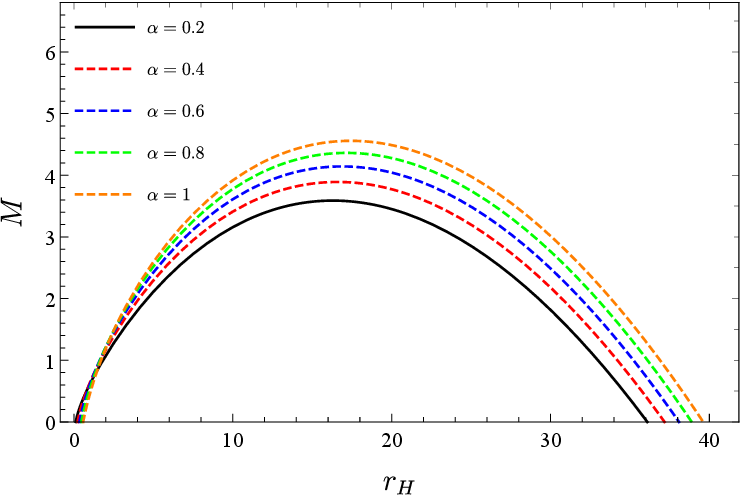}
         \subcaption{ $ \omega_{q}=-0.55$.}\label{fig:Mbb}
   \end{minipage}\ 
\begin{minipage}[t]{0.5\textwidth}
        \centering
        \includegraphics[width=\textwidth]{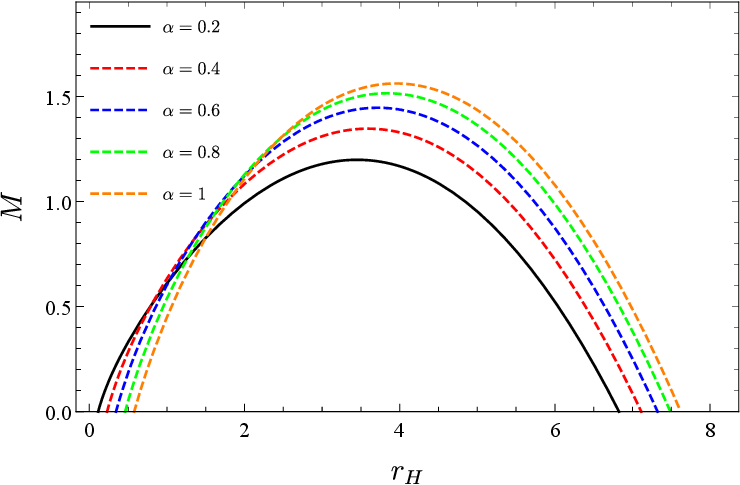}
       \subcaption{ $\omega_{q}=-0.75$.}\label{fig:Mcc}
   \end{minipage}%
\begin{minipage}[t]{0.5\textwidth}
        \centering
        \includegraphics[width=\textwidth]{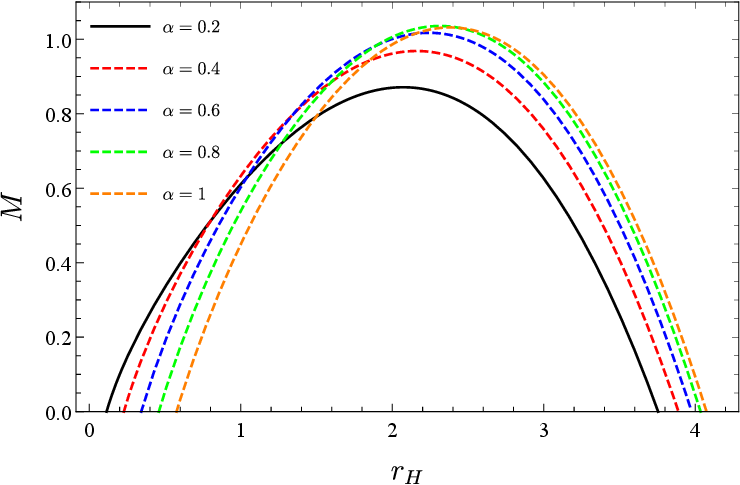}
         \subcaption{ $ \omega_{q}=-0.95$.}\label{fig:Mdd}
   \end{minipage}
\caption{The impact of the dark matter and dark energy on the Schwarzschild black hole mass for $\sigma=0.1 $.}
\label{figmass}
\end{figure}

Fig. \ref{fig:Maa}{, shows that the black hole mass grows as $\alpha$ increases, for the same horizon size, indicating that dark matter may enhance the black hole's mass. The nearly linear behavior of the curves suggests a proportional relationship between mass and horizon size, with a slight deviation at smaller radii that could be due to additional dark matter effects}.
In these cases, we observe that there is a lower limit. In Figs. \ref{fig:Mbb}, \ref{fig:Mcc}, and \ref{fig:Mdd},
{ we see that all curves exhibit a similar parabolic shape, starting near the origin. As $\alpha$ increases, the curves reach higher peak masses, suggesting that a higher dark matter content allows for more massive black holes. After reaching a maximum, the mass begins to decrease until it approaches zero. When the mass approaches its maximum, the black hole may undergo a change in stability, This extremum of the Hawking mass can obtained from $\frac{\partial }{%
\partial r_{H}}M=0.$}.
In all these cases, we observe that there is a lower limit as well as an upper limit on the event horizon. To be more precise, we numerically calculate the lower and upper bound event horizon radii for these scenarios and present them in Table \ref{tab:1}.
\begin{table}[htb!]
    \centering
    \begin{tabular}{c|c c|c c|c c|c c}
    \hline \hline
    \rowcolor{lightgray}\multicolumn{1}{c|}{$\alpha$} & \multicolumn{2}{c|}{$\omega _{q}=-0.35$} & \multicolumn{2}{c|}{$\omega _{q}=-0.55$} & \multicolumn{2}{c|}{$\omega _{q}=-0.75$} & \multicolumn{2}{c}{$\omega _{q}=-0.95$} \\ \hline
 \rowcolor{lightgray} & $r_{H_{min}}$ & $r_{H_{max}}$ & $r_{H_{min}}$ & $r_{H_{max}}$  & $r_{H_{min}}$ & $r_{H_{max}}$ & $r_{H_{min}}$ & $r_{H_{max}}$ \\ \hline
0.2 & 0.117 & $\infty$ & 0.114 & 36.093 & 0.114 & 6.827 & 0.114 & 3.755\\
0.4 & 0.235 & $\infty$ & 0.230 & 37.177 & 0.228 & 7.114 & 0.227 & 3.890\\
0.6 & 0.352 & $\infty$ & 0.347 & 38.087 & 0.344 & 7.325 & 0.342 & 3.976\\
0.8 & 0.470 & $\infty$ & 0.464 & 38.888 & 0.460 & 7.489 & 0.458 & 4.035\\
1.0 & 0.588 & $\infty$ & 0.582 & 39.610 & 0.578 & 7.622 & 0.575 & 4.075\\
\hline\hline
    \end{tabular}
    \caption{The event horizon range that determines physical black hole mass with $\sigma=0.1$. }
    \label{tab:1}
\end{table}
We observe that for smaller quintessence state parameter values, the lower and upper bound horizon radii values take smaller values. We also notice that the dark matter parameter affects the maximum mass value of the black hole in such a way that for greater dark matter parameters, the black hole achieves its maximum mass value at a larger event horizon.
\section{Black Hole's Thermodynamics} \label{sec3}

In this section, we will derive the important thermal functions of the black hole we are considering and discuss the impact of dark matter and dark energy on them. As usual, we start with the Hawking temperature which can be obtained in the semi-classical framework using the relationship between surface gravity and metric components \cite{WHawking}
\begin{equation}
T_{H}=\frac{\kappa }{2\pi }=\frac{1}{4\pi}\frac{d}{dr}f\left( r\right)\bigg\vert_{r=r_H}. \label{tt}
\end{equation}%
Substituting Eqs. (\ref{f}) and (\ref{m11}) in Eq. (\ref{tt}), we obtain a relation between the Hawking temperature and event horizon radius
\begin{equation}
T_{H}=\frac{1}{4\pi r_{H}}\left( 1+\frac{\alpha }{r_{H}}+\frac{3\sigma
\omega _{q}}{r_{H}^{1+3\omega _{q}}}\right) .  \label{t1}
\end{equation}%
We observe that the Hawking temperature, which depends on the parameters of the dark energy and dark matter, reduces to the ordinary case for $\alpha =\sigma =0$ \cite{Vazque}. Moreover, only in the absence of the quintessence field Eq. (\ref{t1}) turns to 
\begin{eqnarray}
 T_{H}=\frac{1}{ 4\pi r_{H}}\left( 1+\frac{\alpha }{r_{H}}\right),   
\end{eqnarray}
which has to be named as the Hawking temperature of the Schwarzschild black hole in the background of PFDM. Similarly, only in the absence of the dark matter Eq. (\ref{t1}) reduces to the Hawking temperature of Schwarzschild black hole surrounded by the dark energy in the form of   \cite{BB1}
\begin{eqnarray}
    T_{H}=\frac{1}{4\pi r_{H}}\left( 1+%
\frac{3\sigma \omega _{q}}{r_{H}^{1+3\omega _{q}}}\right).
\end{eqnarray} 
Now, we examine the Hawking temperature within four cases as above. To this end, we plot Fig. \ref{fig:temp1} and illustrate the variation of Hawking temperatures with respect to the event horizon.
\begin{figure}[tbh]
\begin{minipage}[t]{0.5\textwidth}
        \centering
        \includegraphics[width=\textwidth]{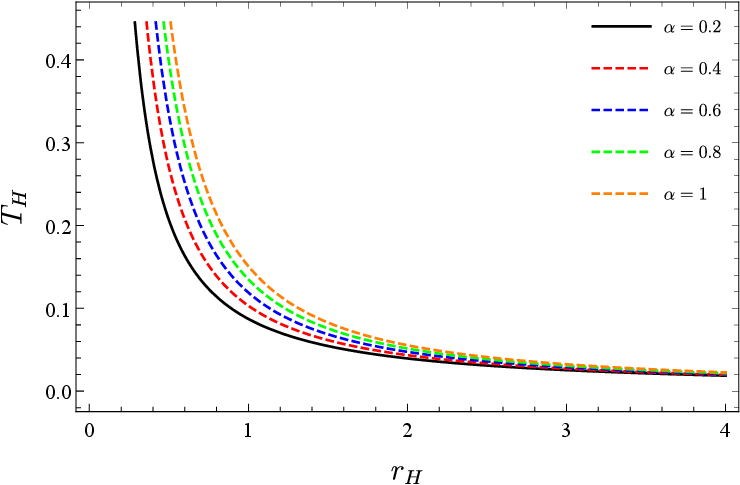}
            \subcaption{ $ \omega_{q}=-0.35$.}\label{fig:Ta}
   \end{minipage}%
\begin{minipage}[t]{0.50\textwidth}
        \centering
       \includegraphics[width=\textwidth]{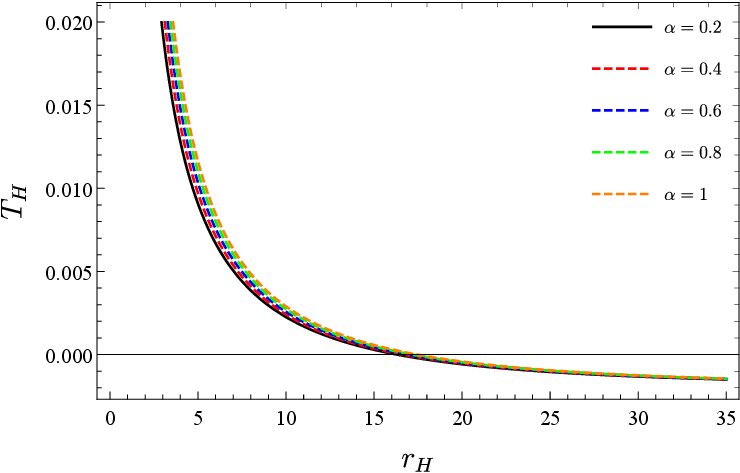}\\
             \subcaption{ $ \omega_{q}=-0.55$.}\label{fig:Tb}
    \end{minipage} \
    \begin{minipage}[t]{0.5\textwidth}
        \centering
        \includegraphics[width=\textwidth]{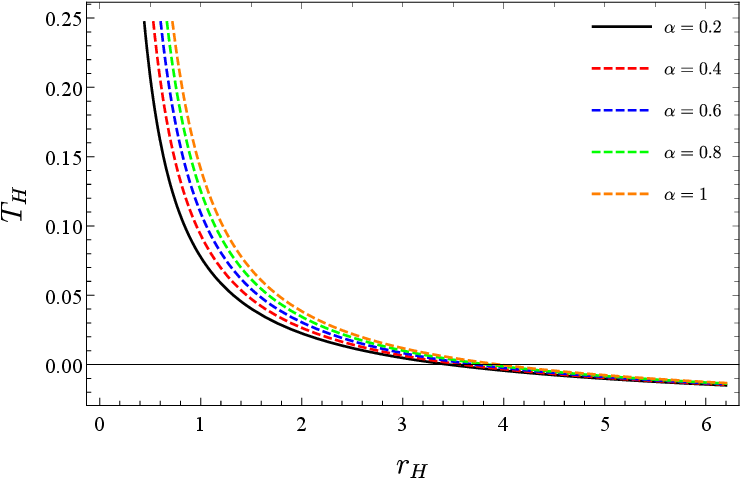}
       \subcaption{ $\omega_{q}=-0.75$.}\label{fig:Tcc}
   \end{minipage}%
\begin{minipage}[t]{0.5\textwidth}
        \centering
        \includegraphics[width=\textwidth]{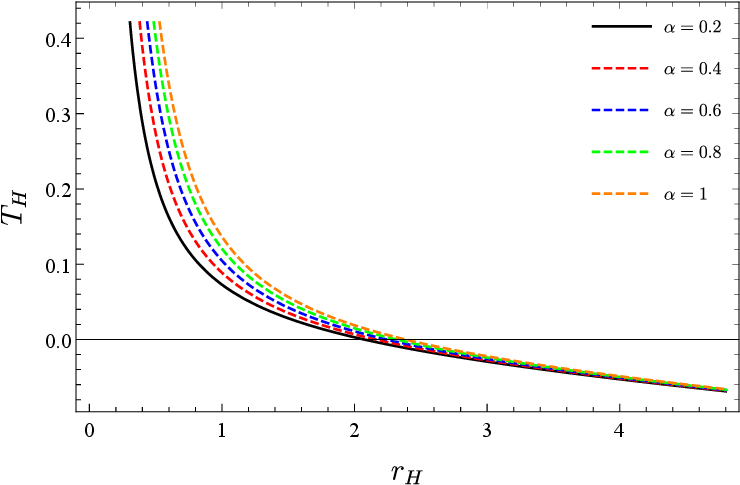}
         \subcaption{ $ \omega_{q}=-0.95$.}\label{fig:Tdd}
   \end{minipage}
    \hfill
\caption{Behavior of the Hawking temperature in terms of event horizon for different values of $\alpha$ and $\sigma =0.1.$}
\label{fig:temp1}
\end{figure}

\newpage
Fig. \ref{fig:temp1} shows that the Hawking temperature decreases in all cases as the event horizon increases similar to the ordinary case. However, only for $\omega_q=-0.35$, the Hawking temperature tends to zero at infinite horizon radius as in the standard case. For the other three cases, {the Hawking temperature $T_{H}$ decreases as the event horizon radius $r_{H}$ increases, eventually approaching zero for very large black holes. This implies that larger black holes become colder and emit less radiation. As $\alpha$  increases, the curves shift upward and rightward,  indicating that the presence of PFDM increases the temperature for black holes of the same size. This effect is more pronounced for smaller black holes, as evidenced by the greater divergence of curves at smaller radii.  However, as $r_{H}$ grows, the temperature converges to zero for all values of $\alpha$, suggesting that the effect of PFDM diminishes for more massive black holes. The behavior of the temperature approaching zero hints at a potential final state where black holes reach a stable, cold configuration, with no further Hawking radiation being emitted. This could suggest the formation of black hole remnants. For the larger values of the event horizon radius the temperature $T_{H}$ becomes negative. Recent experiments \cite{Braun}, have proven the existence of particles with negative temperatures in their motion. In \cite{Norte}, the author studied the possibility of such negative temperatures existing within black holes and proposed that this could generate an outward pressure counteracting the gravitational pull that draws matter inward. }

Then, we explore the effect of the dark energy. To determine it, we depict  Fig. \ref{fig:temp2} using two fixed values of the PFDM parameter. 
\begin{figure}[htb!]
\begin{minipage}[t]{0.5\textwidth}
        \centering
        \includegraphics[width=\textwidth]{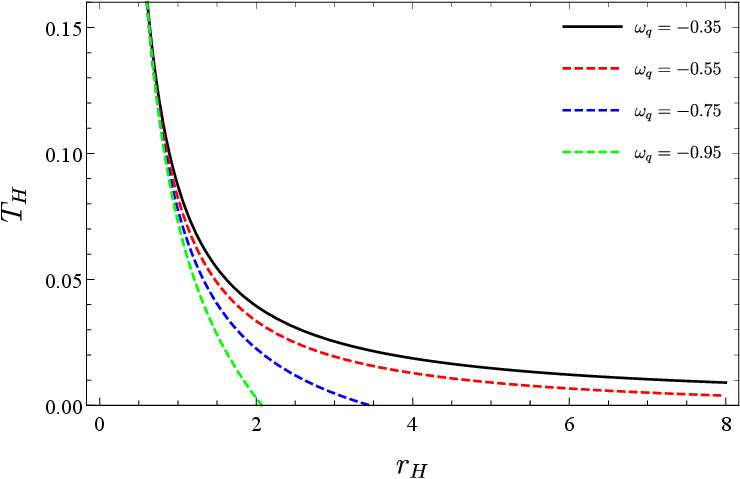}
            \subcaption{ $\alpha=0.2.$}\label{fig:Tc}
   \end{minipage}%
\begin{minipage}[t]{0.5\textwidth}
        \centering
       \includegraphics[width=\textwidth]{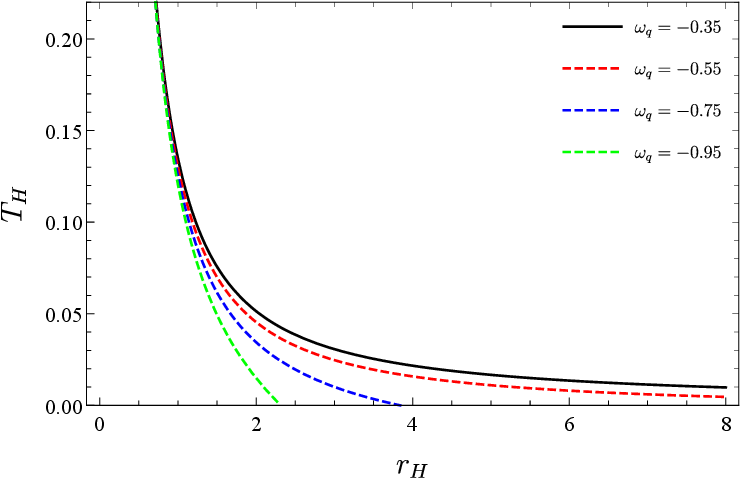}\\
            \subcaption{ $ \alpha=0.8$.}\label{fig:Td}
    \end{minipage}\hfill
\caption{Behavior of the Hawking temperature in terms of event horizon for different values of $\omega _{q}$ and $\sigma =0.1.$}
\label{fig:temp2}
\end{figure}

\newpage
We observe that for greater quintessence state parameters Hawking temperature decreases faster. Now, we use numerical methods to calculate the upper bound of the event horizon and the black hole mass values at those radii. We tabulate our results in Table \ref{tab:2}.
\begin{table}[htb!]
    \centering
    \begin{tabular}{c|c c|c c|c c|c c}
    \hline \hline
    \rowcolor{lightgray}\multicolumn{1}{c|}{$\alpha$} & \multicolumn{2}{c|}{$\omega _{q}=-0.35$} & \multicolumn{2}{c|}{$\omega _{q}=-0.55$} & \multicolumn{2}{c|}{$\omega _{q}=-0.75$} & \multicolumn{2}{c}{$\omega _{q}=-0.95$} \\ \hline
 \rowcolor{lightgray} & $r_H$ & $M_H$ & $r_H$ & $M_H$  & $r_H$ & $M_H$ & $r_H$ & $M_H$ \\ \hline
0.2 & $\infty$ & $\infty$ & 16.294 & 3.589 & 3.450 & 1.199 & 2.072 & 0.871\\
0.4 & $\infty$ & $\infty$ & 16.578 & 3.891 & 3.589 & 1.347 & 2.160 & 0.968\\
0.6 & $\infty$ & $\infty$ & 16.874 & 4.143 & 3.717 & 1.447 & 2.240 & 1.017\\
0.8 & $\infty$ & $\infty$ & 17.152 & 4.362 & 3.838 & 1.515 & 2.314 & 1.036\\
1.0 & $\infty$ & $\infty$ & 17.424 & 4.558 & 3.951 & 1.562 & 2.382 & 1.032\\
\hline\hline
    \end{tabular}
    \caption{The value of the event horizon that makes the Hawking temperature zero. The second column denotes the black hole mass value at that radius.}
    \label{tab:2}
\end{table}

Next, we compute the entropy using the first law of black hole thermodynamics, $S=\int \frac{dM}{T}$. We find 
\begin{equation}
S=\pi r_{H}^{2}=\frac{A}{4}.
\end{equation}%
We note that the PFDM and the dark energy do not affect the entropy of the event horizon which is consistent with the existing findings in the literature \cite{BB1}.

Now, we move on to explore phase transitions and stability. To achieve our purpose, we examine the heat capacity function 
\begin{equation}
C=\frac{dM}{dT},  \label{c}
\end{equation}%
in the presence of PFDM and dark energy. Taking into account Eqs. (\ref{m11}) and (\ref{t1}), Eq. (\ref{c}) gives the heat capacity function in the form of
\begin{equation}
C = - 2\pi r_{H}^{2}\frac{1+\frac{\alpha }{r_{H}}+\frac{3\sigma \omega_{q}}{r_{H}^{1+3\omega _{q}}}} {1+\frac{2\alpha }{r_{H}}+\frac{3\sigma \omega _{q}\left( 3\omega _{q}+2\right) }{r_{H}^{1+3\omega _{q}}}}.
\end{equation}%
It is remarkable that for $\alpha = \sigma = 0$ the capacity function returns to its standard form. 

\newpage
In Fig. \ref{figspeheat} we display the variation of the heat capacity versus the event horizon for the chosen parameters above. 
\begin{figure}[htb!]
\begin{minipage}[t]{0.48\textwidth}
        \centering
        \includegraphics[width=\textwidth]{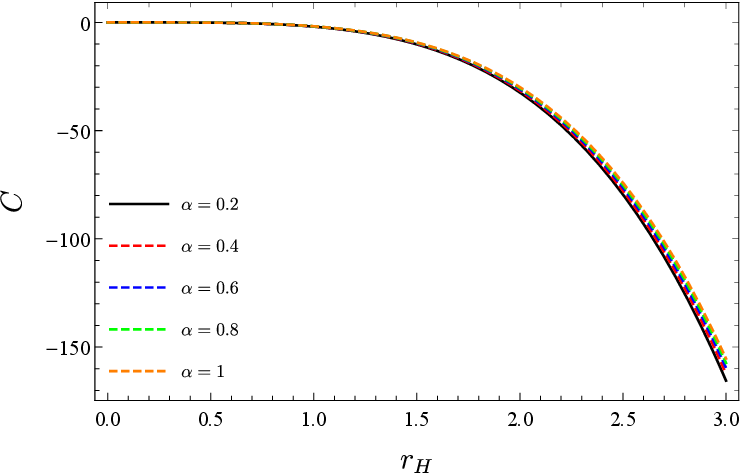}
       \subcaption{ $ \omega_{q}=-0.35$.}\label{fig:Ca1}
   \end{minipage}%
\begin{minipage}[t]{0.5\textwidth}
        \centering
        \includegraphics[width=\textwidth]{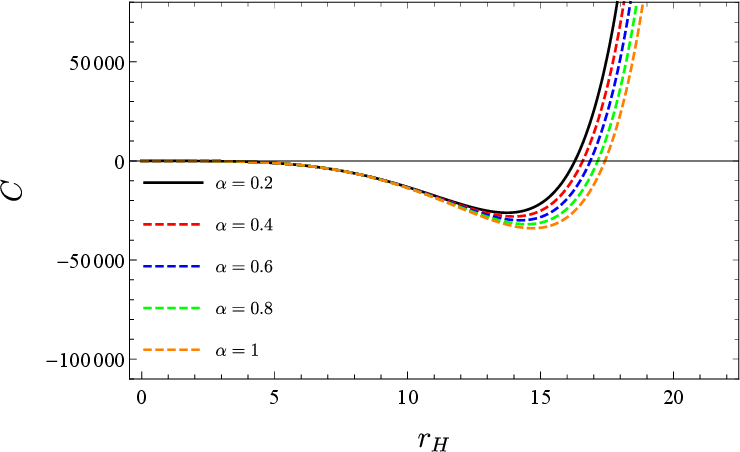}
         \subcaption{ $ \omega_{q}=-0.55$.}\label{fig:Ca2}
   \end{minipage}\ 
\begin{minipage}[t]{0.5\textwidth}
        \centering
        \includegraphics[width=\textwidth]{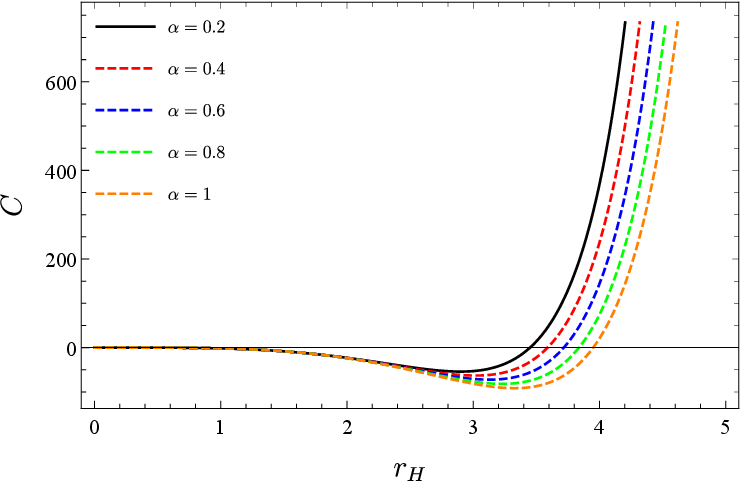}
       \subcaption{ $\omega_{q}=-0.75$.}\label{fig:Ca3}
   \end{minipage}%
\begin{minipage}[t]{0.5\textwidth}
        \centering
        \includegraphics[width=\textwidth]{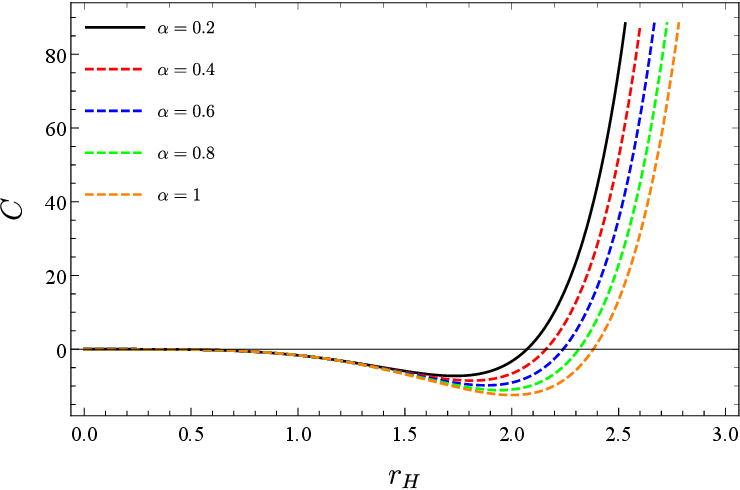}
         \subcaption{ $ \omega_{q}=-0.95$.}\label{fig:Ca4}
   \end{minipage}
\caption{The impact of the dark matter and dark energy on the heat capacity function of Schwarzschild black hole for $\sigma=0.1 $.}
\label{figspeheat}
\end{figure}

As we emphasized before, we see that the $\omega _{q}=-0.35$ scenario differs from the other cases. Fig. \ref{fig:Ca1} shows that { as $r_{H}$ increases, the heat capacity becomes more negative, suggesting larger black holes have a greater tendency towards instability \cite{HawPage1983}. Additionally, smaller values of the PFDM parameter $\alpha$ correspond to more negative heat capacities, implying that lesser dark matter presence leads to more pronounced black hole instability}.
 In other cases, we observe that the heat capacity is first negative and then positive. This indicates a first-order phase transition at the horizon where specific heat equals zero. In other words, the vanishing value of the specific heat states that the black hole ceases to exchange radiation with its surroundings. Therefore, at that horizon black hole becomes stable and a remnant mass occurs. To clarify these discussions, we calculate the event horizon radii where specific heat disappears and present their values in Table \ref{tab:3}.  
\begin{table}[htb!]
    \centering
    \begin{tabular}{c|c|c|c|c}
    \hline \hline
    \rowcolor{lightgray} $\alpha$ & {$\omega _{q}=-0.35$} & {$\omega _{q}=-0.55$} & {$\omega _{q}=-0.75$} & {$\omega _{q}=-0.95$} \\ \hline
  \rowcolor{lightgray} & $r_{H}$ & $r_{H}$ & $r_{H}$ & $r_{H}$   \\ \hline
0.2 & 0 &  16.294 &  3.450 &  2.072 \\
0.4 & 0 &  16.578 &  3.589 &  2.160 \\
0.6 & 0 &  16.874 &  3.717 &  2.240 \\
0.8 & 0 &  17.152 &  3.838 &  2.314 \\
1.0 & 0 &  17.424 &  3.951 &  2.382 \\
\hline\hline
    \end{tabular}
    \caption{Event horizon values where the heat capacity function equals zero.}
    \label{tab:3}
\end{table}

\newpage
We obtain two important results here. The first of these is that as the dark matter parameter increases, the event horizon radii also increase. Secondly, and more importantly, we see that the horizon values exactly coincide with those given in Table \ref{tab:2}. This means that the mass values in Table \ref{tab:2} are the remnant masses.

Next, we examine the equation of state. To this end, we employ the relation between pressure and $\sigma$ parameter 
\begin{equation}
P=-\frac{\sigma }{8\pi },  \label{g1}
\end{equation}%
and rewrite the Hawking temperature in terms of pressure
\begin{equation}
P=-\frac{r_{H}^{1+3\omega _{q}}}{24\pi \omega _{q}}\bigg(1+\frac{\alpha }{r_{H}}-{4\pi r_{H}T_{H}}\bigg).  \label{pr}
\end{equation}
In Fig. \ref{figper}, we display the pressure isotherms in terms of the horizon.
\begin{figure}[htb!]
\begin{minipage}[t]{0.5\textwidth}
        \centering
        \includegraphics[width=\textwidth]{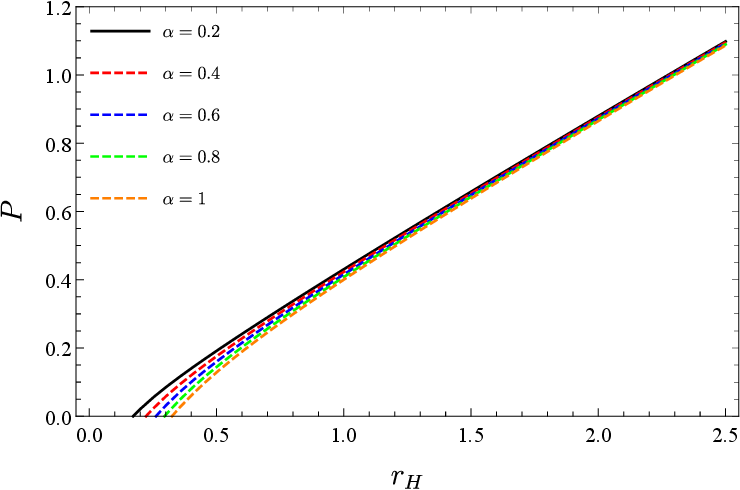}
            \subcaption{ $ \omega_{q}=-0.35$.}\label{fig:pa}
   \end{minipage}
\begin{minipage}[t]{0.5\textwidth}
        \centering
        \includegraphics[width=\textwidth]{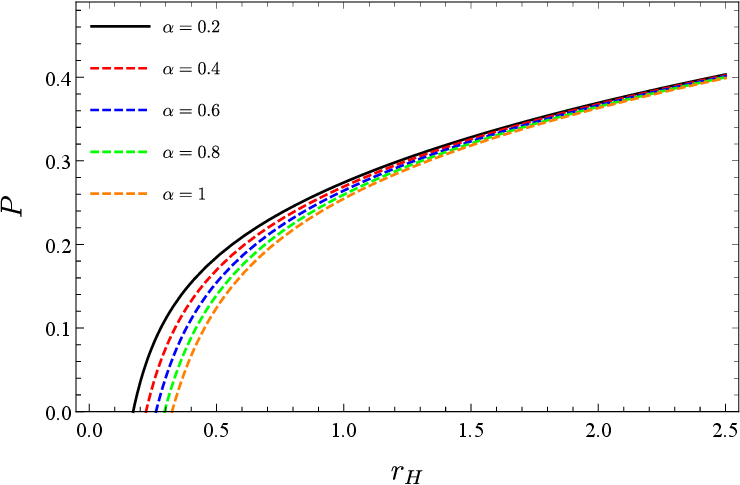}
            \subcaption{ $ \omega_{q}=-0.55$.}\label{fig:pb}
   \end{minipage}\hfill   
\begin{minipage}[t]{0.5\textwidth}
        \centering
        \includegraphics[width=\textwidth]{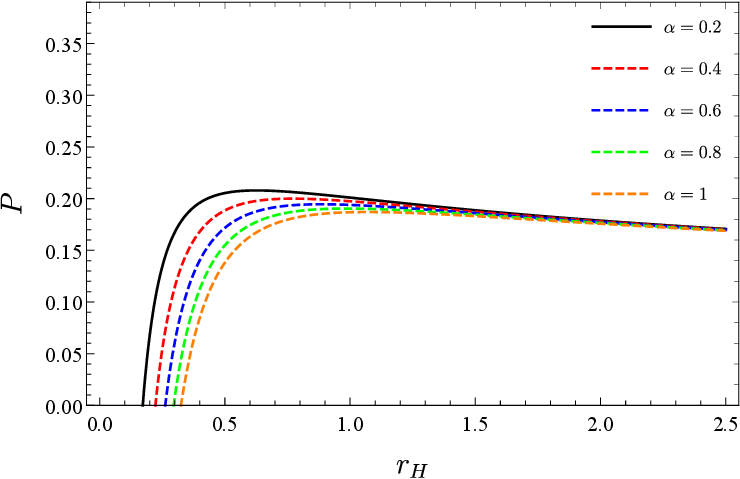}
            \subcaption{ $ \omega_{q}=-0.75$.}\label{fig:pc}
   \end{minipage}  
\begin{minipage}[t]{0.50\textwidth}
        \centering
       \includegraphics[width=\textwidth]{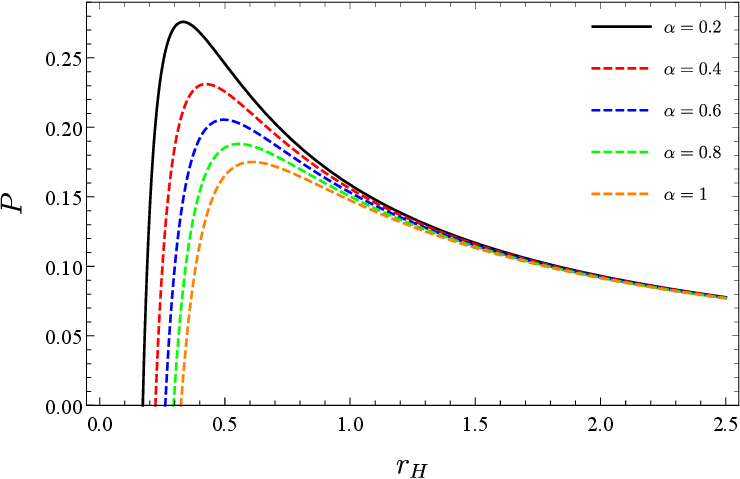}\\
            \subcaption{ $\omega_{q}=-0.95$.}\label{fig:pd}
    \end{minipage}\hfill
\caption{The impact of the dark matter on the pressure isotherms of $T=1$.}
\label{figper}
\end{figure}

\newpage
We observe that dark matter alters the equation of state at relatively smaller horizons. this effect becomes significant for smaller quintessence state parameters. 
For example, in the case $\omega_{q}=-0.95$, the peak of the equation of state nearly disappears for greater dark matter parameters. 

\section{Black Hole's Shadow} \label{sec4}

In this section, we calculate the shape of the shadow cast of the black hole we are considering and investigate the effect of dark matter and quintessence field on these forms. To achieve our goal, we will derive the equation of motion using the Lagrangian formulation. Then, we will employ the Hamilton-Jacobi equation to study the geodesic equation near the black hole. We start with the Lagrangian,
\begin{equation}
\mathcal{L}=\frac{1}{2}g_{\mu \nu }\dot{x}^{\mu }\dot{x}^{\nu },
\end{equation}%
where we use the dot notation over the variable to indicate differentiation with respect to the proper time. Here, for the static spacetime metric with spherical symmetry, we consider the Lagrangian of the form  
\begin{equation}
\mathcal{L}=\frac{1}{2}\left[ -f\left( r\right) \dot{t}^{2}+\frac{1}{f\left(
r\right) }\dot{r}^{2}+r^{2}\dot{\theta}^{2}+r^{2}\sin ^{2}\theta \dot{\varphi%
}^{2}\right],
\end{equation}
Using the Euler-Lagrange equation
\begin{equation}
\frac{\partial \mathcal{L}}{\partial x^{\mu }}-\frac{d}{d\tau }\frac{%
\partial \mathcal{L}}{\partial \dot{x}^{\mu }}=0,
\end{equation}
we get two conserved quantities
\begin{equation}
\left( 1-\frac{2M}{r}-\frac{\sigma }{r^{3\omega _{q}+1}}+\frac{\alpha }{r}%
\ln \frac{r}{\left\vert \alpha \right\vert }\right) \text{ }\dot{t}=E,
\end{equation}%
\begin{equation}
r^{2}\sin ^{2}\theta \text{ }\dot{\varphi}=L,
\end{equation}%
where $E$ is the energy of the test particle and $L$ is the angular momentum. Then, we analyze the orbits of photons around the black hole with the help of the Hamilton-Jacobi equation
\begin{equation}
\frac{\partial }{\partial \tau }\mathcal{S}=-\frac{1}{2}g_{\mu \nu }\frac{%
\partial \mathcal{S}}{\partial x^{\mu }}\frac{\partial \mathcal{S}}{\partial
x^{\nu }},  \label{21}
\end{equation}%
where the Jacobi action is denoted with $\mathcal{S}$. We then form the geodesic equations via Carter's approach \cite{Carter1968}
\begin{equation}
\mathcal{S}=\frac{1}{2}m^{2}\tau -Et+\mathcal{S}_{r}\left( r\right) +%
\mathcal{S}_{\theta }\left( \theta \right) +L\varphi ,  \label{22}
\end{equation}%
Using this separation of variables we acquire the equation of motion of photons 
\begin{equation}
-\frac{E^{2}}{f\left( r\right) }+f\left( r\right) \left( \frac{\partial 
\mathcal{S}_{r}}{\partial r}\right) ^{2}+\frac{1}{r^{2}}\left( \frac{%
\partial \mathcal{S}_{\theta }}{\partial \theta }\right) ^{2}+\frac{L^{2}}{%
r^{2}\sin ^{2}\theta }=0.  \label{s13}
\end{equation}%
Now, we employ the relationships $P_{\theta }=\frac{\partial \mathcal{L}}{\partial 
\dot{\theta}}=\frac{\partial }{\partial \theta }\mathcal{S}_{\theta }$  and 
$P_{r}=\frac{\partial \mathcal{L}}{\partial 
\dot{r}}=\frac{\partial }{\partial r}\mathcal{S}_{r}$, and we obtain 
\begin{eqnarray}
\frac{\partial }{\partial \theta }\mathcal{S}_{\theta }&=&r^{2}\dot{\theta},
\label{s11}  \\
\frac{\partial }{\partial r}\mathcal{S}_{r}&=&\frac{1}{f\left( r\right) }\dot{r%
}.  \label{s12}
\end{eqnarray}%
Substituting Eqs. (\ref{s11}) and (\ref{s12}) in Eq. (\ref{s13}), we find two separated equations with the following form 
\begin{eqnarray}
r^{4}\dot{r}^{2}&=&r^{4}E^{2}-r^{2}\left( 1-\frac{2M}{r}-\frac{\sigma }{%
r^{3\omega _{q}+1}}+\frac{\alpha }{r}\ln \frac{r}{\left\vert \alpha
\right\vert }\right) \left( \mathcal{K}+L^{2}\right) ,\,\,\,\,   \label{255} \\
r^{4}\dot{\theta}^{2}&=&\mathcal{K}-L^{2}\cot ^{2}\theta .  \label{28}
\end{eqnarray}%
Here, $\mathcal{K}$ is Carter's the separation constant. We then express the complete null geodesic equation
\begin{eqnarray}
\dot{t}&=&\frac{E}{f\left( r\right) },  \label{e} \\
\dot{\varphi}&=&\frac{L}{r^{2}\sin ^{2}\theta},  \label{p} \\
r^{2}\dot{r}&=&\pm \sqrt{\mathcal{R}},  \label{r} \\
r^{2}\dot{\theta}&=&\pm \sqrt{\Theta},  \label{th}
\end{eqnarray}
where
\begin{eqnarray}
\mathcal{R}&=&r^{4}E^{2}-r^{2}\left( \mathcal{K}+L^{2}\right) \left( 1-\frac{2M%
}{r}-\frac{\sigma }{r^{3\omega _{q}+1}}+\frac{\alpha }{r}\ln \frac{r}{%
\left\vert \alpha \right\vert }\right) ,\,\,\,\, \\
\Theta&=&\mathcal{K}-L^{2}\cot ^{2}\theta .
\end{eqnarray}
We observe that two impact parameters characterize photon motion in the vicinity of the black hole
\begin{eqnarray}
   \eta &=& \frac{L}{E}, \\
   \zeta ^{2}&=&\frac{\mathcal{K}}{E^{2}},
\end{eqnarray}
which depends on the constants of motion and Carter's separation constant. These impact parameters govern key properties of the photon's trajectory influenced by
the gravitational field of the black hole. 

We now focus on the unstable circular orbits of photons that encode the boundaries of shadow casts. To find these boundaries, we revisit the equation that describes the radial motion of photons 
\begin{equation}
\dot{r}^{2}+V_{eff}\left( r\right) =0,  \label{vpot}
\end{equation}
with the effective potential of the radial motion
\begin{equation}
V_{eff}\left( r\right) =\left( \frac{\mathcal{K}+L^{2}}{r^{2}}\right) \left(
1-\frac{2M}{r}-\frac{\sigma }{r^{3\omega _{q}+1}}+\frac{\alpha }{r}\ln \frac{%
r}{\left\vert \alpha \right\vert }\right) -E^{2}.
\end{equation}
Due to three possible trajectories of photons approaching the black hole, namely falling into it, scattering away, or forming a circular orbit near the black hole, the shape of the shadow is influenced primarily by the unstable circular orbits. We can identify these orbits by imposing the following conditions
\begin{equation}
\left. V_{eff}\left( r\right) \right\vert _{r=r_{ph}}=0,  \label{vpot1}
\end{equation}%
and 
\begin{equation}
\left. \frac{d}{dr}V_{eff}\left( r\right) \right\vert _{r=r_{ph}}=0,
\label{vpot2}
\end{equation}%
with 
\begin{eqnarray}
    \left. \frac{d^{2}}{dr^{2}}V_{eff}\left( r\right) \right\vert_{r=r_{p}}<0,
\end{eqnarray}
which indicates a maxima for the effective potential at $r=r_{ph}$. At first, we use Eq. (\ref{vpot1}) and find
\begin{equation}
\eta +\zeta ^{2}=\frac{r_{ph}^{2}}{1-\frac{2M}{r_{ph}}-\frac{\sigma }{%
r_{ph}^{3\omega _{q}+1}}+\frac{\alpha }{r_{ph}}\ln \frac{r_{ph}}{\left\vert
\alpha \right\vert }}.
\end{equation}%
Then, we utilize  Eq. (\ref{vpot2}), and obtain a relationship 
\begin{equation}
r_{ph}f^{\prime }\left( r_{ph}\right) -2f\left( r_{ph}\right) =0,
\label{psr}
\end{equation}
in terms of $r_{ph}$. After we substitute the {metric} function, Eq. (\ref{f}), we arrive at a subsequent equation 
\begin{equation}
2-\frac{6M}{r_{ph}}-\frac{\sigma \left( 3\omega _{q}+3\right) }{%
r_{ph}^{3\omega _{q}+1}}+\frac{3\alpha }{r_{ph}}\ln \frac{r_{ph}}{\left\vert
\alpha \right\vert }-\frac{\alpha }{r_{ph}}=0.
\end{equation}%
This equation describing the photon sphere has no analytical solution, so we have to resort to numerical methods to reach the solution. We tabulate the calculated photon sphere radii and the impact parameter of the photon sphere, $\eta +\zeta^{2}$, for different values of dark matter and quintessence field state parameters in  Table \ref{tabmm}. 
\newpage
\begin{table}[htb!]
\centering
\begin{tabular}{l|ll|ll|ll|ll}
\hline \hline
\rowcolor{lightgray}\multirow{2}{*}{} & \multicolumn{2}{c|}{$\omega _{q}=-0.35$} & 
\multicolumn{2}{c|}{$\omega _{q}=-0.55$} & \multicolumn{2}{c|}{$\omega _{q}=-0.75$} & \multicolumn{2}{c}{$\omega _{q}=-0.95$} \\ \hline
\rowcolor{lightgray}
$\alpha $ & \multicolumn{1}{l}{$r_{p}$} & $\eta +\zeta ^{2}$ &  \multicolumn{1}{l}{$r_{p}$} & $\eta +\zeta ^{2}$ & \multicolumn{1}{l}{$r_{p}$} & $\eta +\zeta ^{2}$ & \multicolumn{1}{l}{$r_{p}$} & $\eta +\zeta ^{2}$ \\ \hline
$0.2$ & $2.59645$ & $20.9170$ & $2.66310$ & $20.9574$ & $2.66306$ & $20.9573$ & $2.44482$ & 
$21.1831$ \\ 
$0.4$ & $2.37328$ & $15.9573$ & $2.41058$ & $15.9696$ & $2.39378$ & $15.9611$ & $2.24088$
& $16.1484$ \\ 
$0.6$ & $2.31904$ & $14.0390$ & $2.34807$ & $14.0461$ & $2.33044$ & $14.0401$ & $2.20121$ & 
$14.1805$ \\ 
$0.8$ & $2.34719$ & $13.4236$ & $2.37539$ & $13.4300$ & $2.36056$ & $13.4251$ & $2.23941$ & 
$13.5350$ \\ \hline\hline
\end{tabular}
\label{tabm1}
\caption{{}The values of the photon radius, $r_{p}$, and $\protect\eta +\zeta ^{2}$ for different values of $\protect\alpha $ and $\omega _{q}$ with $\sigma =0.1$ and $M=1$. }
\label{tabmm}
\end{table}
We now introduce Celestial coordinates, $X$ and $Y$, to describe the actual shadow of the black hole seen in an observer's frame \cite{Vazque}  
\begin{eqnarray}
X&=&\lim_{r_{o}\rightarrow \infty }\left( -r_{o}^{2}\sin \theta _{o}\frac{d\varphi }{dr}\right) ,  \label{XX} \\
Y&=&\lim_{r_{o}\rightarrow \infty }\left( r_{o}^{2}\frac{d\theta }{dr}\right). \label{YY}
\end{eqnarray}%
Here, $r_{o}$ is the location of the observer, and $\theta _{o}$ is the angle of inclination. Using the geodesics equations, we simplify Eqs. (\ref{XX}) and (\ref{YY}) to a simpler form
\begin{eqnarray}
X&=&-\frac{\zeta }{\sin \theta _{0}}, \\
Y&=&\sqrt{\eta -\zeta ^{2}\cot ^{2}\theta _{0}} .
\end{eqnarray}
The preceding two equations establish a connection between the celestial coordinates and the constants of motion. We then consider the equatorial plane with $\theta _{0}=\pi /2$, so the celestial coordinates simplify to 
\begin{equation}
X^{2}+Y^{2}=\eta +\zeta ^{2}=R_{S}^{2}.
\end{equation}%
Here, $R_{S}$ is the radius of the shadow. 

Now, we display the effect of the PFDM and quintessence matter field on the black hole shadows. First, in Fig. \ref{fig:shad1} we present the impact of dark matter with four distinguished dark energy scenarios. We observe that 
for smaller dark matter parameters the shadows have greater radii.

\newpage
\begin{figure}[htb!]
\begin{minipage}[t]{0.50\textwidth}
        \centering
        \includegraphics[width=\textwidth]{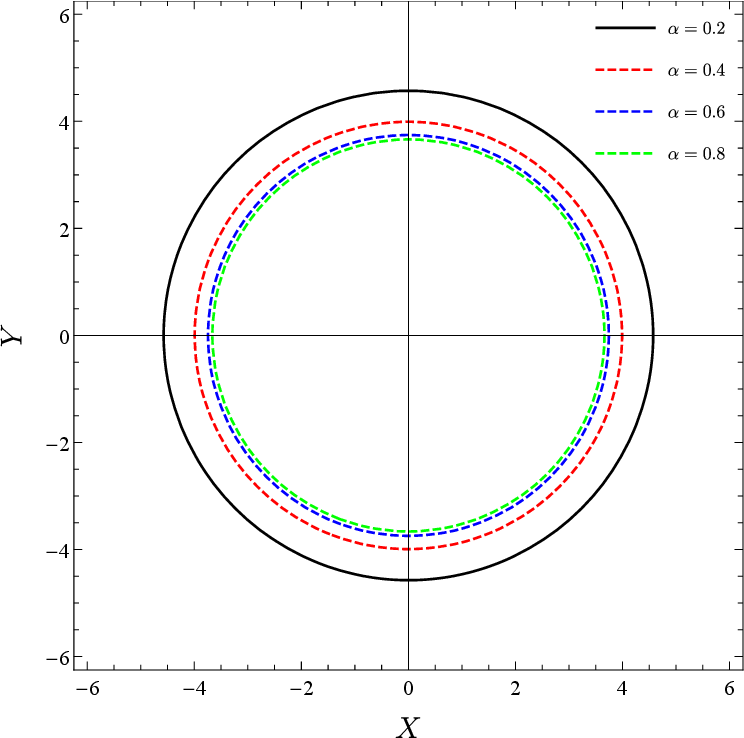}
       \subcaption{ $\omega_{q}=-0.35$.}\label{fig:Spa}
   \end{minipage}%
\begin{minipage}[t]{0.50\textwidth}
        \centering
        \includegraphics[width=\textwidth]{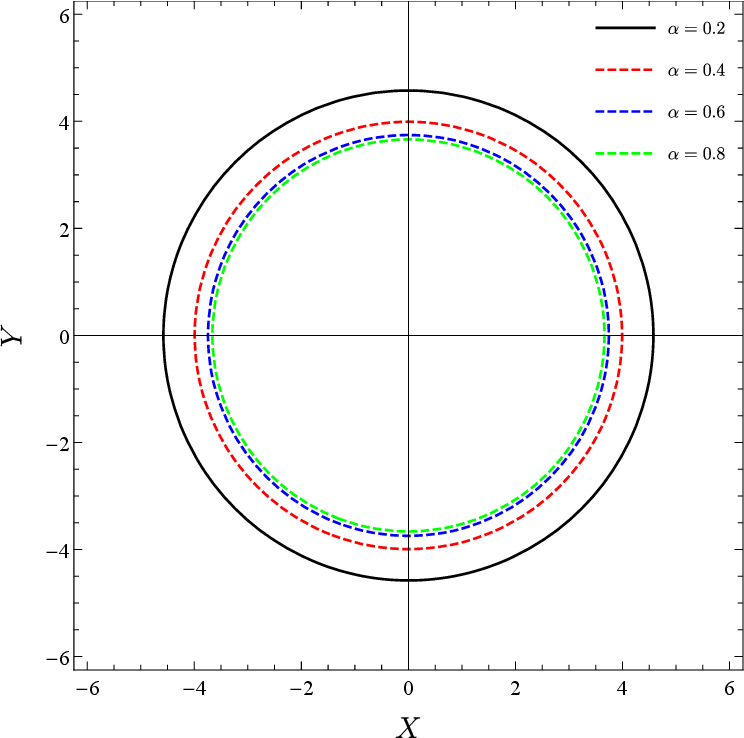}
         \subcaption{ $\omega_{q}=-0.55$.}\label{fig:Spb}
   \end{minipage}\ 
\begin{minipage}[t]{0.50\textwidth}
       \centering
        \includegraphics[width=\textwidth]{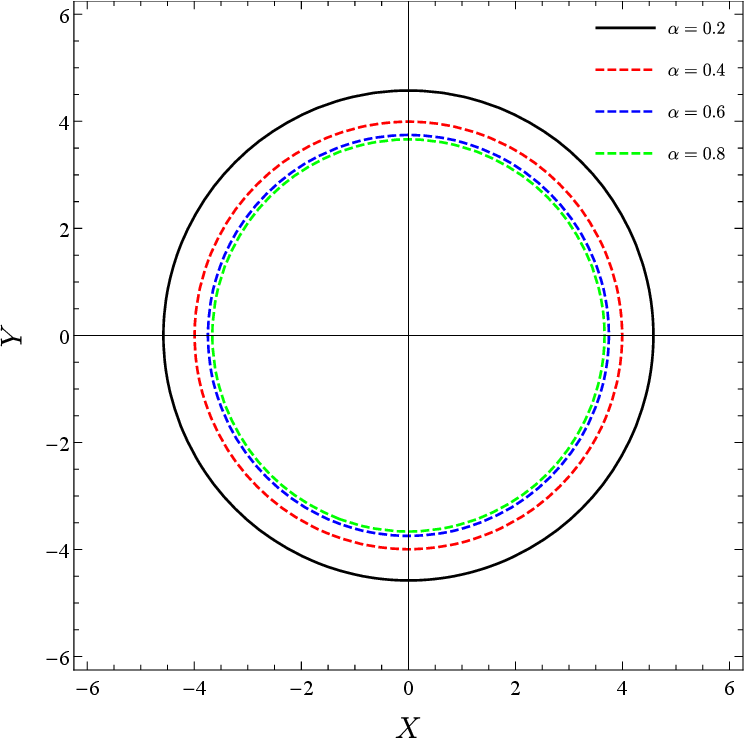}
       \subcaption{$\omega_{q}=-0.75$.}\label{fig:Spc}
   \end{minipage}%
\begin{minipage}[t]{0.50\textwidth}
        \centering
        \includegraphics[width=\textwidth]{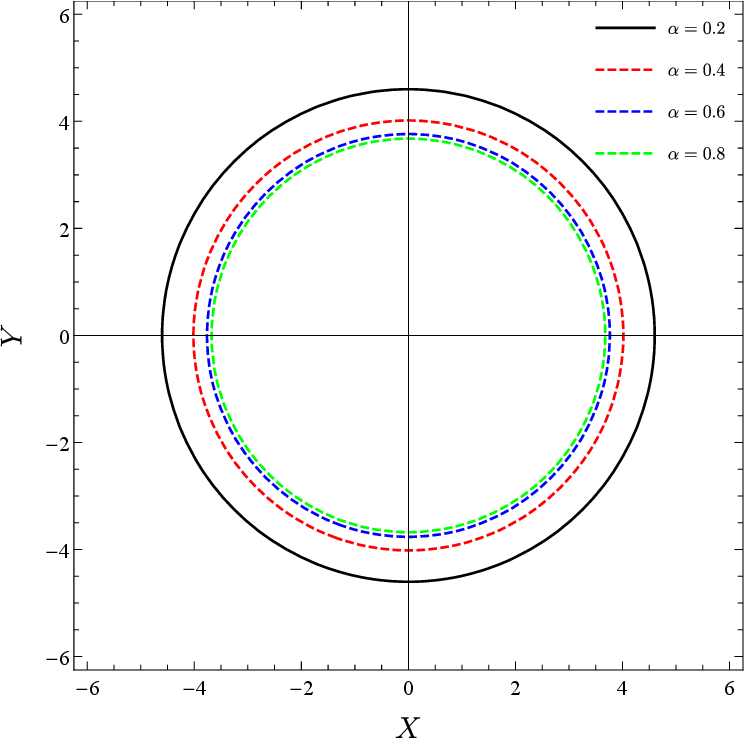}
       \subcaption{ $\omega_{q}=-0.95$.}\label{fig:Spd}
   \end{minipage}
\caption{The impact of the dark matter on the shadows of the Schwarzschild black with  $M=1$ and $\sigma$=0.1.} 
\label{fig:shad1}
\end{figure}
However, in these graphs, we cannot clarify the effect of the quintessence matter field in a fixed dark matter scenario. To explain this, we plot Fig. \ref{fig:shad2}.

\newpage
\begin{figure}[htb!]
\begin{minipage}[t]{0.50\textwidth}
        \centering
        \includegraphics[width=\textwidth]{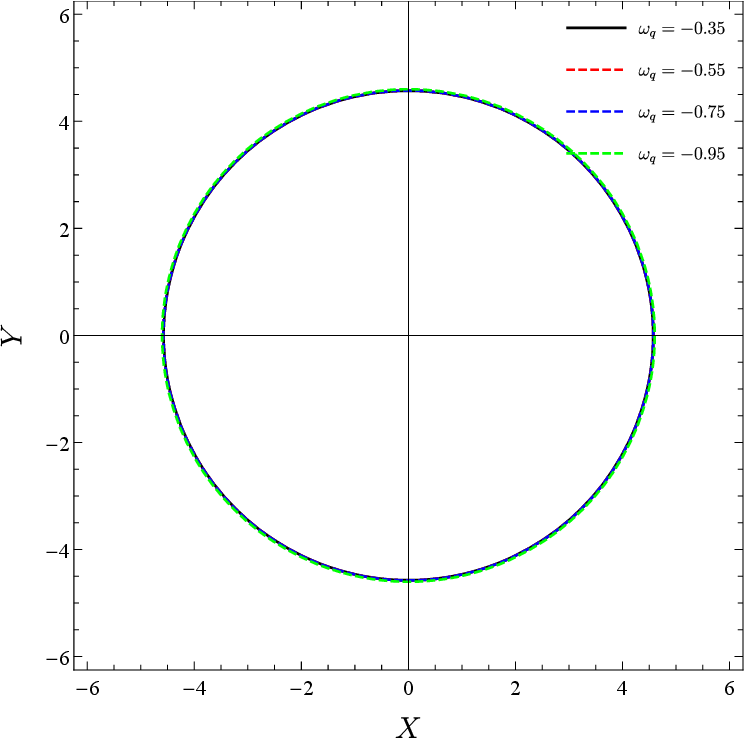}
       \subcaption{ $ \alpha=0.2$.}\label{fig:Spaa}
   \end{minipage}%
\begin{minipage}[t]{0.50\textwidth}
        \centering
        \includegraphics[width=\textwidth]{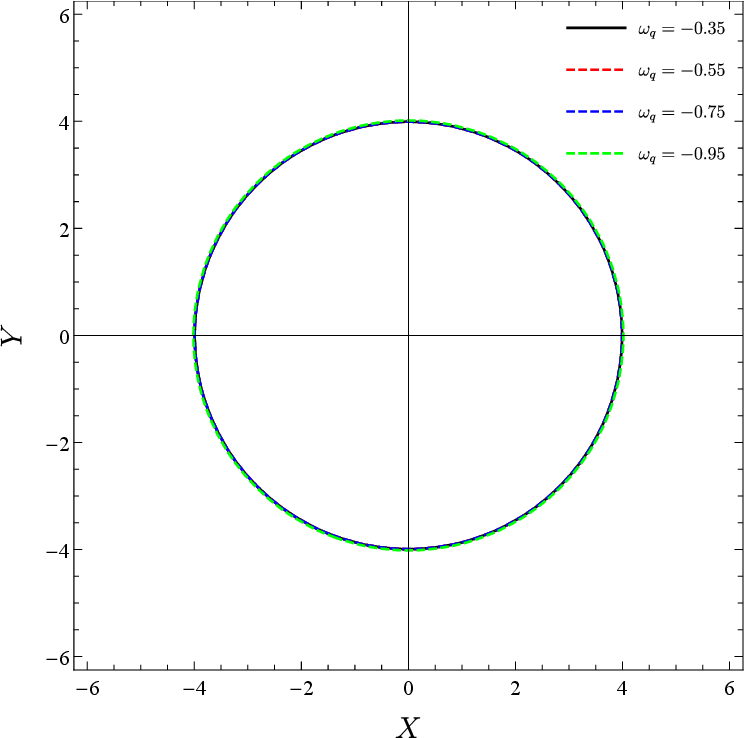}
         \subcaption{ $\alpha=0.4$.}\label{fig:Spbb}
   \end{minipage}\ 
\begin{minipage}[t]{0.50\textwidth}
       \centering
        \includegraphics[width=\textwidth]{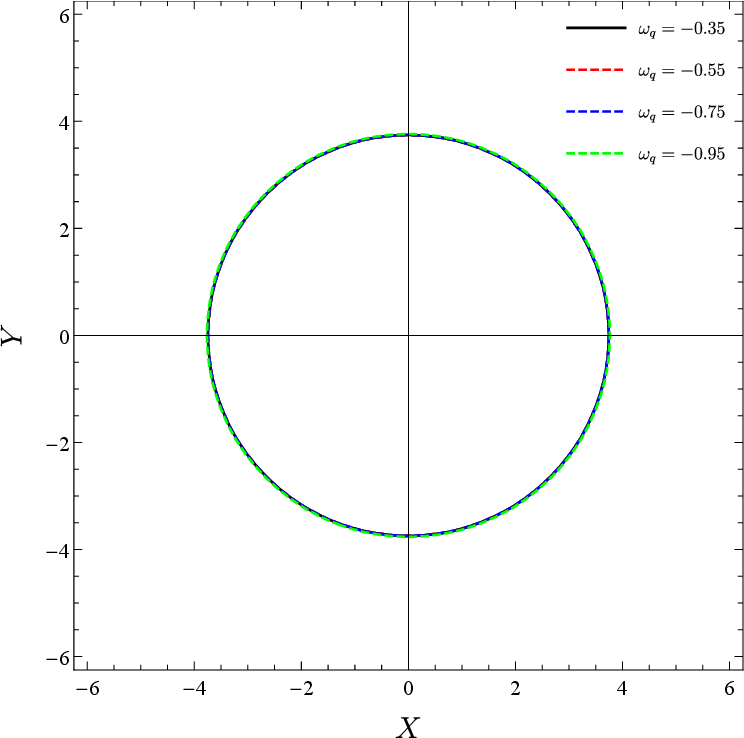}
       \subcaption{  $\alpha=0.6$.}\label{fig:Spcc}
   \end{minipage}%
\begin{minipage}[t]{0.50\textwidth}
        \centering
        \includegraphics[width=\textwidth]{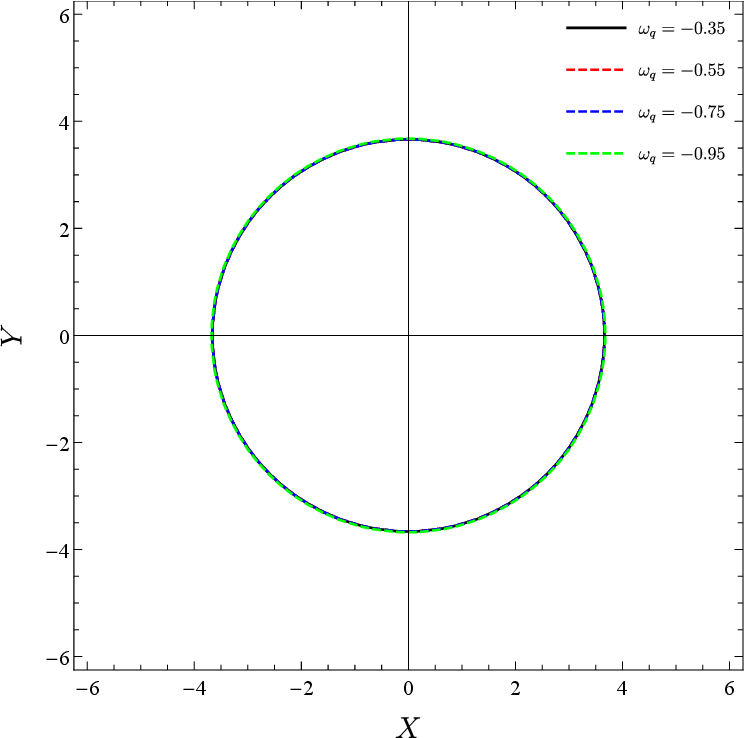}
       \subcaption{ $\alpha=0.8$.}\label{fig:Spdd}
   \end{minipage}
\caption{The impact of the dark energy on the shadows of the Schwarzschild black with  $M=1$ and $\sigma$=0.1.}
\label{fig:shad2}
\end{figure}
We observe that the influence of the quintessence matter field on the shadow radius is limited. We revisit these effects in Fig. \ref{fig:shad3}.

\newpage
\begin{figure}[htb!]
\begin{minipage}[t]{0.5\textwidth}
        \centering
        \includegraphics[width=\textwidth]{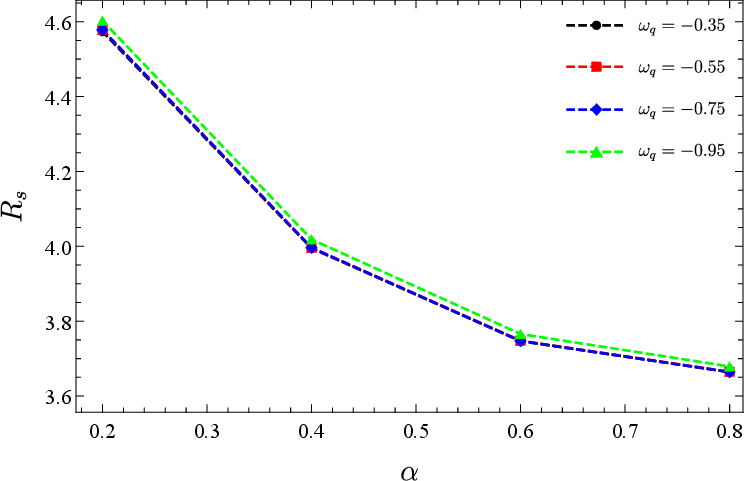}
      \subcaption{ } \label{fig:shad3a}
   \end{minipage}%
\begin{minipage}[t]{0.50\textwidth}
        \centering
       \includegraphics[width=\textwidth]{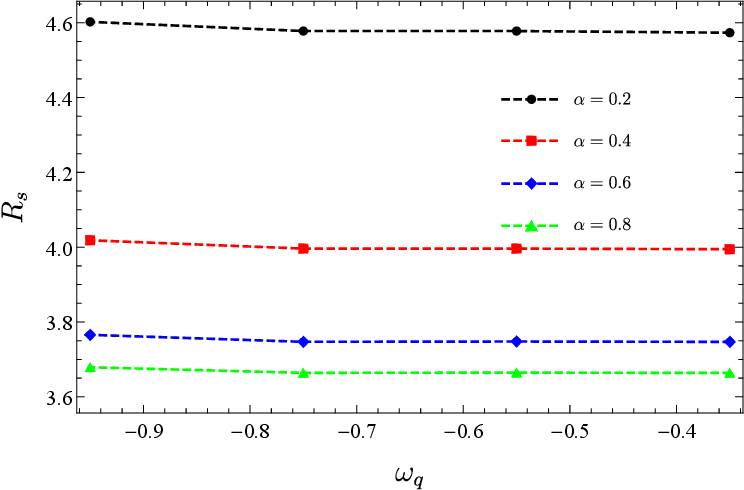}\\
             \subcaption{ } \label{fig:shad3b}
    \end{minipage}\hfill
\caption{The variation of the radius of the black hole shadow  forfixed $M$ and $\sigma $ values.}
\label{fig:shad3}
\end{figure}

Fig. \ref{fig:shad3a} displays that an increase in the dark matter parameter leads to a decrease in the shadow radius, We observe that this impact is the same for all quintessence matter cases. Fig. \ref{fig:shad3b} shows that the shadow radius slightly decreases as the quintessence matter increases for a fixed dark matter parameter.

\section{Black Hole's Quasinormal Modes} \label{sec5}
After a significant event such as the merger of black holes, the ringdown phase exhibits an intriguing behavior called quasinormal modes. These modes display distinct oscillation patterns not influenced by the initial disturbances that caused them. In other words, the quasinormal modes arise from the intrinsic properties of the system itself, representing the natural vibrations of spacetime, irrespective of the precise initial conditions that triggered them. They gradually emit energy through gravitational waves and they are associated with open systems, as opposed to normal modes, which are linked to closed systems.

In the literature, we observe distinct methodologies for the calculation of their frequencies \cite{Mod1, Cho2003, Jing2004, Mod4, Mod3, Juricb}. One of the most preferred ones requires these modes by solving the wave equation within the context of a system defined by a background metric. However, deriving analytical solutions for these modes is often an extremely challenging endeavor, and various approaches are used to obtain them. Among these, the Wentzel-Kramers-Brillouin (WKB) approach stands out with its widespread use.

\subsection{WKB approach}
The WKB was initially proposed in \cite{Schutz, Mashhoon} at the first order, and subsequently extended to higher orders by Iyer, Opala, and Konoplya  \cite{Iyer, Konoplya, Opala}. The WKB approach gained popularity in recent decades because, unlike more advanced numerical methods, it automatically applies to diverse effective potentials with sufficient accuracy.

Our purpose here is to consider the massless scalar quasinormal mode in the Schwarzschild spacetime surrounded by quintessence in the background of PFDM. We start with the  Klein--Gordon differential equation of the curved spacetime
\begin{equation}
\frac{1}{\sqrt{-g}}\partial _{\mu }\left( \sqrt{-g}g^{\mu \nu }\partial
_{\nu }\psi \right) =0,  \label{q1}
\end{equation}
to describe the propagation of the massless scalar field, $\psi$. By taking the spherical symmetry into account, we decompose the scalar field in a specific manner 
\begin{equation}
\psi \left( t,r,\theta ,\varphi \right) =e^{-i\omega t}\frac{\Psi _{\omega
,L}\left( r\right) }{r}Y_{L,\mu }\left( \theta ,\varphi \right) ,
\label{dec}
\end{equation}
where $\omega $ and $Y_{L,\mu }\left( \theta ,\varphi\right) $ denote the frequency and  the spherical harmonics, respectively. Then, we substitute this decomposition in Eq. (\ref{q1}), and get a Schrodinger-like equation that exhibits wave-like properties, 
\begin{equation}
\frac{d^{2}}{dr^{\ast 2}}\Psi \left( r^{\ast }\right) -\left( \omega
^{2}-V\left( r^{\ast }\right) \right) \Psi \left( r^{\ast }\right) =0. \label{eqto}
\end{equation}
Here, we use the tortoise variable 
\begin{equation}
dr^{\ast }=\frac{dr}{f\left(r\right) },    
\end{equation}
with the Regge-Wheeler potential
\begin{equation}
V\left( r\right) =\frac{f\left( r\right) }{r}\frac{df\left( r\right) }{dr}+%
\frac{f\left( r\right) L\left( L+1\right) }{r}.
\end{equation}%
Now, we utilize Eq. \ref{f} and depict the Regge-Wheeler potential with different dark matter and quintessence state 
parameter values in Fig. \ref{veff1}. 
\begin{figure}[htb!]
\begin{minipage}[t]{0.5\textwidth}
        \centering
        \includegraphics[width=\textwidth]{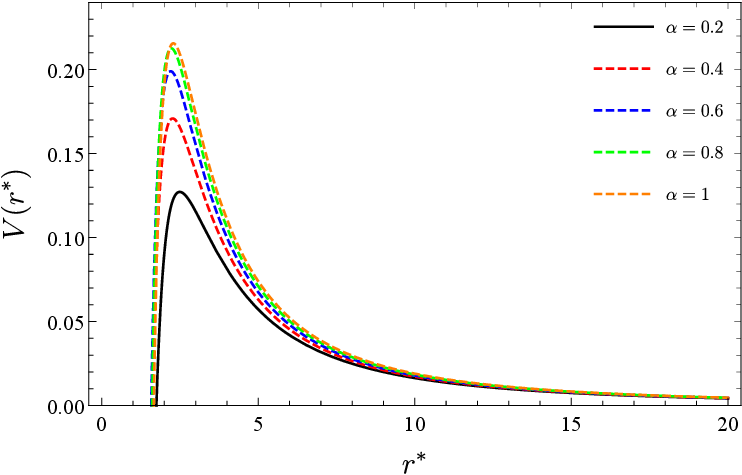}
            \subcaption{ $ \omega_{q}=-0.35$.}\label{figva}
   \end{minipage}%
\begin{minipage}[t]{0.50\textwidth}
        \centering
       \includegraphics[width=\textwidth]{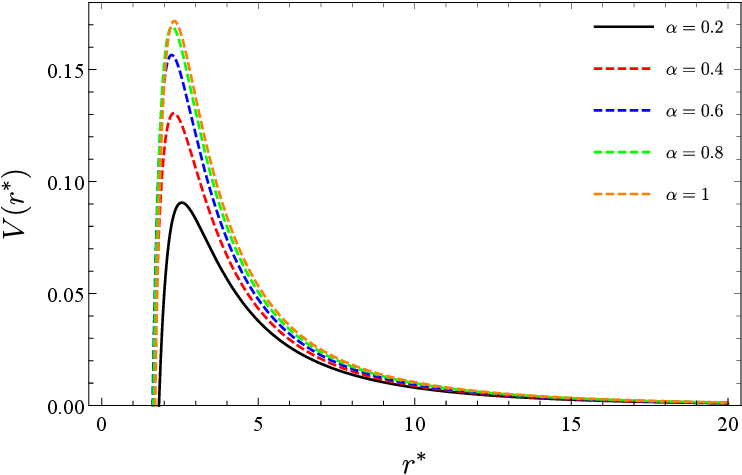}\\
             \subcaption{ $ \omega_{q}=-0.55$.}\label{figvb}
    \end{minipage}\hfill
\begin{minipage}[t]{0.5\textwidth}
        \centering
        \includegraphics[width=\textwidth]{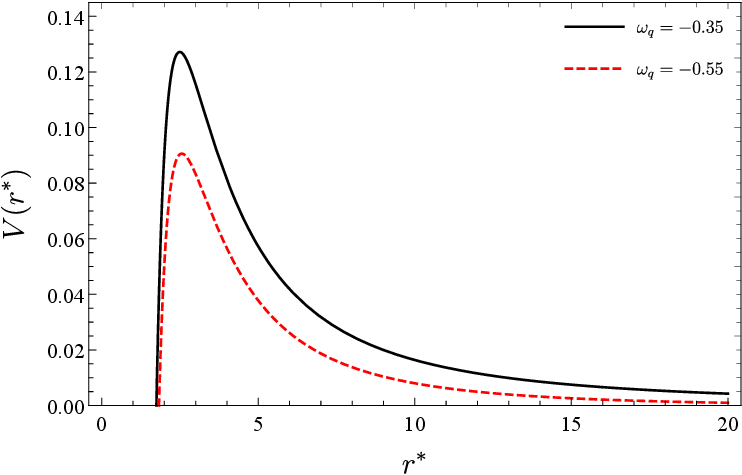}
            \subcaption{ $\alpha=0.2.$}\label{figvc}
   \end{minipage}%
\begin{minipage}[t]{0.5\textwidth}
        \centering
       \includegraphics[width=\textwidth]{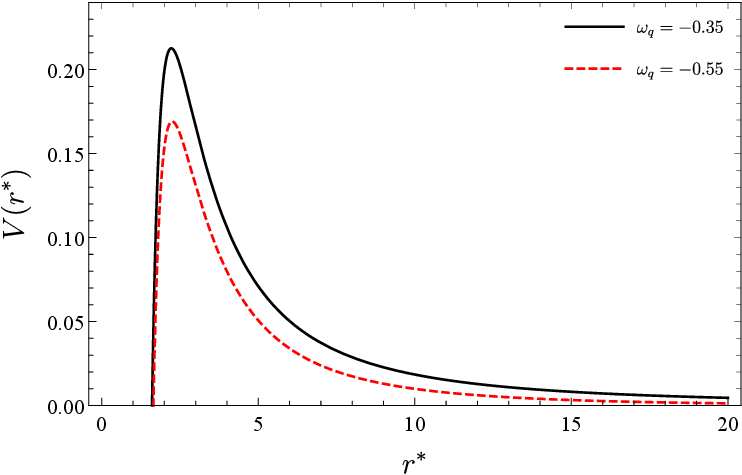}\\
            \subcaption{ $ \alpha=0.8$.}\label{figvd}
    \end{minipage}\hfill
 \caption{Regge-Wheeler potential in terms of $r^{\ast }$ for different values of $\alpha $ and $\omega _{q} $ with $M=1$ and $\sigma =0.1.$}
\label{veff1}
\end{figure}

\newpage
Figs. \ref{figva} and \ref{figvb} show that the peak of the potential becomes significant as the dark matter parameter increases. However, this behavior does not notably alter in the distinct scenarios of quintessence matter fields as indicated in Figs.  \ref{figvc} and \ref{figvd}.

Now, we solve Eq. (\ref{eqto})  under the specified boundary
condition 
\begin{equation}
\Psi \simeq e^{\pm i\omega r^{\ast }}\text{ \ \ \ , }r^{\ast }\rightarrow
\pm \infty ,  \label{eqto1}
\end{equation}
for determining the quasinormal modes. Here. we have to emphasize that the WKB approach can be valid if and only if the potential
has a barrier-like shape and tends to constant values as $r^{\ast
}\rightarrow \pm \infty $. After fitting the power series solution near the potential's turning points, we get the quasinormal modes generator in the form of  \cite{Konoplya}
\begin{equation}
i\frac{\left( \omega _{n}-V_{0}\right) }{\sqrt{-2V_{0}"}}-\sum_{i=2}^{N}%
\Lambda _{i}=n+\frac{1}{2}.  \label{qnmf}
\end{equation}
Here, $n$ is the overtone numbers that take only integer values.  $V_{0}$ is the effective potential height and  $V_{0}"$ is the second derivative to the tortoise coordinate of the potential at its maximum $r_{0}^{\ast }$. $\Lambda _{i}$ is a constant coefficient resulting from higher-order WKB corrections and its explicit expressions were given in \cite{Konoplya, Opala} for the higher orders. Then, with the sixth-order WKB method we calculate the quasinormal frequencies and 
present them in Table \ref{tab:test11}. 
\begin{table}[htb!]
\centering%
\begin{tabular}{l|l|l|l|l|l}
\hline \hline
\rowcolor{lightgray}\multicolumn{5}{c}{$\alpha =0.2$} &  \\ \hline
\rowcolor{lightgray} $L$ & $n$ & $\omega _{q}=-0.35$ & $\omega _{q}=-0.55$ & 
$\omega _{q}=-0.75$ & $\omega _{q}=-0.95$ \\ \hline
1 & 0 & 0.286697-0.130367 i & 0.243049-0.102204 i & 0.167207-0.062235 i & 
0.088169-0.217382 i \\ \hline
2 & 0 & 0.522059-0.116583 i & 0.452545-0.094832 i & 0.291393-0.058030 i & 
0.082422-0.367061 i \\ 
& 1 & 0.472045-0.383570 i & 0.419863-0.313977 i & 0.301527-0.178605 i & 
0.242275-0.408652 i \\ \hline
3 & 0 & 0.748478-0.113292 i & 0.648638-0.092489 i & 0.414522-0.056541 i & 
0.079722-0.519477 i \\ 
& 1 & 0.740130-0.369166 i & 0.642165-0.300088 i & 0.417317-0.174583 i & 
0.241083-0.543687 i \\ 
& 2 & 0.617352-0.665079 i & 0.562526-0.546217 i & 0.434059-0.296269 i & 
0.397733-0.598888 i \\ \hline
4 & 0 & 0.971106-0.111910 i & 0.841437-0.091495 i & 0.536739-0.055883 i & 
0.078399-0.671671 i \\ 
& 1 & 0.963634-0.356097 i & 0.832598-0.292034 i & 0.536914-0.171503 i & 
0.238350-0.687398 i \\ 
& 2 & 0.933763-0.644846 i & 0.815571-0.521053 i & 0.545181-0.292983 i & 
0.399276-0.726612 i \\ 
& 3 & 0.734400-0.931059 i & 0.683972-0.773060 i & 0.566755-0.414447 i & 
0.553285-0.789731 i \\ \hline
\rowcolor{lightgray}\multicolumn{5}{c}{$\alpha =0.5$} &  \\ \hline
1 & 0 & 0.310951-0.146376 i & 0.282983-0.124096 i & 0.251959-0.114566 i & 
0.013906-0.0315182 i \\ \hline
2 & 0 & 0.612428-0.150235 i & 0.552881-0.129377 i & 0.442171-0.102615 i & 
0.013290-0.0535657 i \\ 
& 1 & 0.511360-0.479501 i & 0.474381-0.420532 i & 0.416412-0.448855 i & 
0.038497-0.0587866 i \\ \hline
3 & 0 & 0.886533-0.147911 i & 0.799762-0.127860 i & 0.630862-0.099104 i & 
0.012929-0.0759021 i \\ 
& 1 & 0.863738-0.488408 i & 0.782596-0.421004 i & 0.629265-0.336559 i & 
0.038831-0.0789541 i \\ 
& 2 & 0.641399-0.799494 i & 0.605289-0.708823 i & 0.492779-0.596203 i & 
0.063048-0.0857387 i \\ \hline
4 & 0 & 1.153270-0.146388 i & 1.040150-0.126681 i & 0.817836-0.097655 i & 
0.0127326-0.098205 i \\ 
& 1 & 1.141960-0.471496 i & 1.030190-0.406636 i & 0.813113-0.311049 i & 
0.0386125-0.100088 i \\ 
& 2 & 1.058440-0.854819 i & 0.969550-0.735938 i & 0.794382-0.576247 i & 
0.0640501-0.105080 i \\ 
& 3 & 0.752816-1.083220 i & 0.715545-0.969873 i & 0.576455-0.836921 i & 
0.0874828-0.112720 i \\ \hline\hline
\end{tabular}%
\caption{The quasinormal modes for the massless scalar perturbation  with $M=1$, and $\sigma =0.1$ via the Pade averaged
6th order WKB approximation method.}
\label{tab:test11}
\end{table}

In Fig. \ref{qnmc1} and Fig. \ref{qnmc2} we compare the real and imaginary parts of the quasinormal modes.  

\newpage
\begin{figure}[htb!]
\begin{minipage}[t]{0.5\textwidth}
        \centering
        \includegraphics[width=\textwidth]{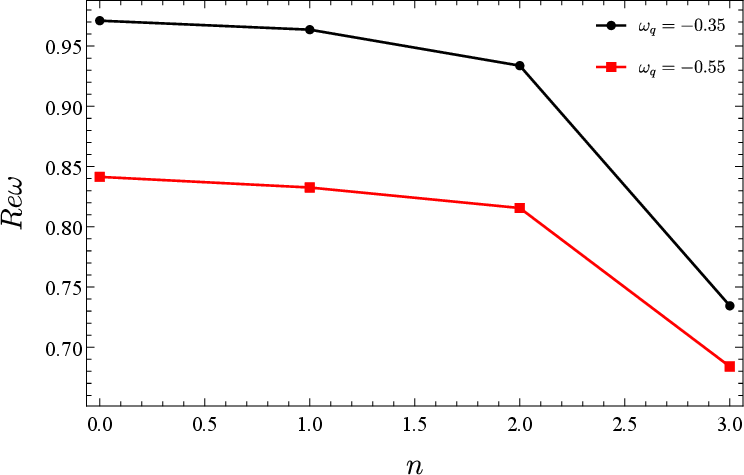}
   \label{fig:qna}
   \end{minipage}%
\begin{minipage}[t]{0.5\textwidth}
        \centering
        \includegraphics[width=\textwidth]{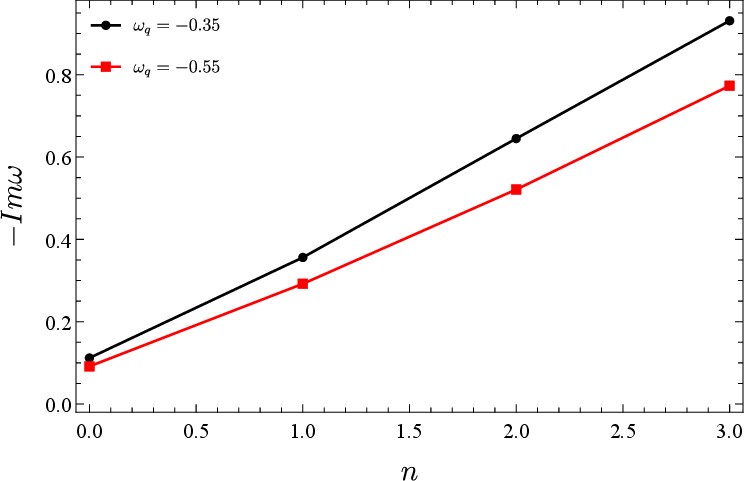}
         \label{fig:qnb}
   \end{minipage}
\caption{The variation of the $\text{Re}\,\omega $ and $\text{Im}\,\omega $ versus overtone numbers for two values of $\omega _{q}=-0.35$, $\omega _{q}=-0.55$, with $M=1$, $\protect\alpha =0.2$, $L=4$.} \label{qnmc1}
\end{figure}

\begin{figure}[htb!]
\begin{minipage}[t]{0.5\textwidth}
      \centering
        \includegraphics[width=\textwidth]{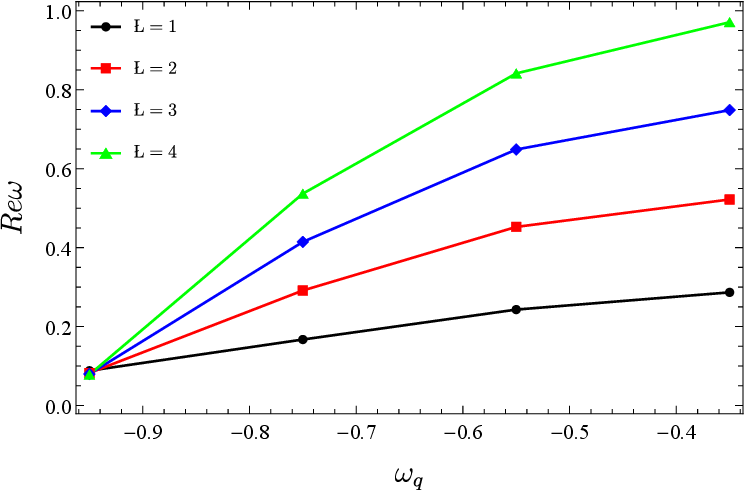}
       \label{fip:qnc}
   \end{minipage}%
\begin{minipage}[t]{0.5\textwidth}
        \centering
        \includegraphics[width=\textwidth]{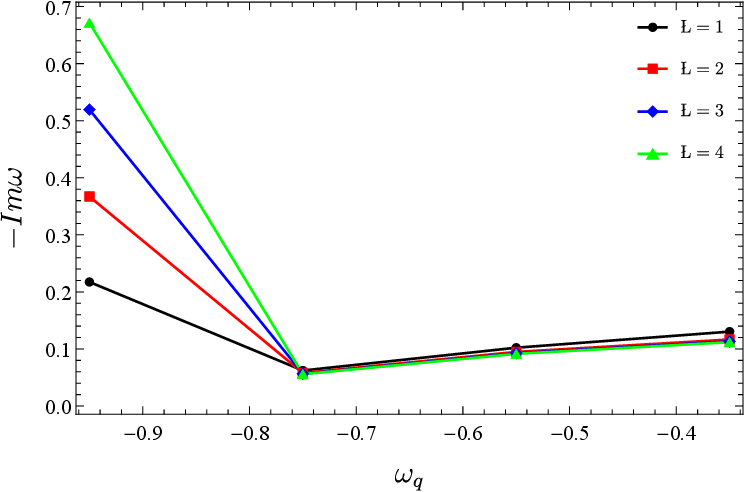}
      \label{fip:qnd}
   \end{minipage}
\caption{The variation of the $\text{Re}\, \omega $ and $\text{Im}\,\omega $ vs. $\protect\omega _{q}$ for $n=0$, $\protect\alpha =0.2$, 
$\protect\sigma =0.1$ and $M=1$. } \label{qnmc2}
\end{figure}

\newpage
We note that the real section of frequencies decreases for greater $n$, while the imaginary section increases. We observe that the real section of quasinormal modes increases at greater values of the quintessence state parameter. Moreover, for greater $L$ this enlargement expands. However, in the imaginary sector, we observe a decrease at first, and then the quasinormal modes increase for greater values of the quintessence state parameter.

The importance of the WKB order becomes evident in determining accurate
values of QNM frequencies. The quantity for $\omega _{k}$, derived through
the WKB formula of order "$k$" for each overtone $n$, is defined as%
\begin{equation}
\Delta =\frac{\left\vert \omega _{k-1}-\omega _{k+1}\right\vert }{2}.
\end{equation}
We compute the quasinormal frequencies and the estimated errors for $L=1$ and  $n=0$. We present the results in Table \ref{taberr1}.
\begin{table}[htbp]
\centering%
\begin{tabular}{l|ll|ll}
\hline\hline
\rowcolor{lightgray}\multirow{2}{*}{} & \multicolumn{2}{c}{$\alpha =0.2$}
& \multicolumn{2}{|c}{$\alpha =0.4$} \\ \hline
\rowcolor{lightgray} $k$ & $\omega _{k}$ & $\Delta _{k}$ & $\omega _{k}$ & $%
\Delta _{k}$ \\ \hline
$3$ & 0.115987-0.0457716 i & 0.00179603 & 0.202929-0.0897261 i & 0.00269626
\\ 
$4$ & 0.11512-0.04104750 i & 0.00256651 & 0.207087-0.0829896 i & 0.00494956
\\ 
$5$ & 0.114298-0.0423931 i & 0.00121871 & 0.198699-0.0839376 i & 0.00418022
\\ 
$6$ & 0.116721-0.0426669 i & 0.00079614 & 0.205075-0.0852250 i & 0.00164852
\\ 
$7$ & 0.115464-0.0423223 i & 0.00059041 & 0.204147-0.0838090 i & 0.00165888
\\ 
$8$ & 0.115525-0.0428855 i & 0.00100064 & 0.203483-0.0855179 i & 0.00139755
\\ 
$9$ & 0.116151-0.0422390 i & 0.00047425 & 0.204282-0.0841058 i & 0.00115333
\\ 
$10$ & 0.115810-0.0426108 i & 0.00030129 & 0.203774-0.0838807 i & 0.00084838
\\ 
$11$ & 0.115923-0.0427487 i & 0.00078338 & 0.203652-0.0841582 i & 
0.00033406 \\ 
$12$ & 0.115919-0.0425271 i & 0.00015844 & 0.203824-0.0838608 i & 
0.00079516 \\ 
$13$ & 0.115959-0.0426456 i & 0.00034563 & 0.203382-0.0838610 i & 0.00081246
\\ \hline\hline
\end{tabular}
\label{tabm2}
\caption{Quasinormal modes and error estimations for $M=1$, $\sigma =0.1$, $\omega _{q}=-0.80$, $L=1$, and $n=0$ through different orders of the WKB approach.}
\label{taberr1}
\end{table}

\newpage
\subsection{Mashhoon approximation}

In  \cite{Ferrari}, Mashhoon proposed to use {P\"{o}schl}-Teller (PT) potential for calculating the quasinormal modes   
\begin{equation}
\mathcal{V}\left( r^{\ast }\right) \sim V_{PT}\left( r^{\ast }\right) =\frac{%
V_{0}}{\cosh ^{2}\beta \left( r^{\ast }-r_{0}^{\ast }\right) },
\end{equation}%
where%
\begin{equation}
\ \beta =\left. \sqrt{ - \frac{1}{2V_{0}}\frac{d^{2}\mathcal{V}\left( r^{\ast
}\right) }{dr^{\ast 2}}}\right\vert _{r^{\ast }=r_{0}^{\ast }}.
\end{equation}
Here, $r_{0}^{\ast }$ is the position of the maximum of $\mathcal{V}\left( r^{\ast
}\right) $ where  $\mathcal{V}\left( r_{0}^{\ast }\right)=V_{0} $. The quasinormal modes of the {PT} potential can be computed analytically
\begin{equation}
\omega _{n}=\pm \beta \sqrt{\frac{V_{0}}{\beta ^{2}}-\frac{1}{4}}-i\beta
\left( n+\frac{1}{2}\right) .  \label{Mas}
\end{equation}
We apply this solution to calculate the quasinormal modes with different dark matter parameter values. We tabulate the results for the scalar field in Table \ref{tabComparison1}. Moreover, we compare them with the frequencies obtained through the 3rd-order WKB method.
\begin{table}[htb!]
\centering%
\begin{tabular}{l|ll|ll}
\hline
\rowcolor{lightgray}\multirow{2}{*}{} & \multicolumn{2}{c}{$\omega _{q}=-0.40$} & 
\multicolumn{2}{c}{$\omega _{q}=-0.60$} \\ \hline
\rowcolor{lightgray}$\alpha $ & WKB method & Mashhoon approximation & WKB method & Mashhoon
approximation \\ \hline
$0.3$ & 0.329678-0.117361 i & 0.356702-0.125127 i & 0.284959-0.095995 i & 
0.297511-0.103264 i \\ 
$0.4$ & 0.348387-0.128153 i & 0.378995-0.137513 i & 0.306336-0.107298 i & 
0.322028-0.116113 i \\ 
$0.5$ & 0.361048-0.137034 i & 0.395447-0.148050 i & 0.320987-0.116472 i & 
0.339968-0.126847 i \\ 
$0.6$ & 0.368699-0.144095 i & 0.406926-0.156764 i & 0.330112-0.123658 i & 
0.352384-0.135553 i \\ 
$0.7$ & 0.372336-0.149506 i & 0.414282-0.163773 i & 0.334837-0.129067 i & 
0.360273-0.142399 i \\ 
$0.8$ & 0.372844-0.153473 i & 0.418299-0.169246 i & 0.336148-0.132941 i & 
0.364530-0.147595 i \\ \hline\hline
\end{tabular}%
\caption{The comparison of the quasinormal modes estimated with the Mashhoon approximation and 3rd order WKB method, for specific parameters: $M=1,$ $L=1, $ $n=0,$ $\protect\sigma =0.1$, and varying values of the parameter $\alpha $.}
\label{tabComparison1}
\end{table}

Finally, we present a graphical comparison in Fig. \ref{qnmn}. 
\begin{figure}[htb!]
\begin{minipage}[t]{0.5\textwidth}
        \centering
        \includegraphics[width=\textwidth]{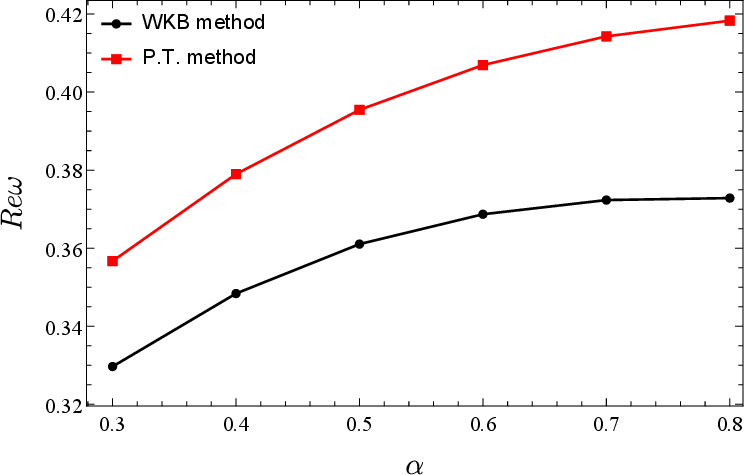}
       \label{fig:qnna}
   \end{minipage}%
\begin{minipage}[t]{0.5\textwidth}
        \centering
        \includegraphics[width=\textwidth]{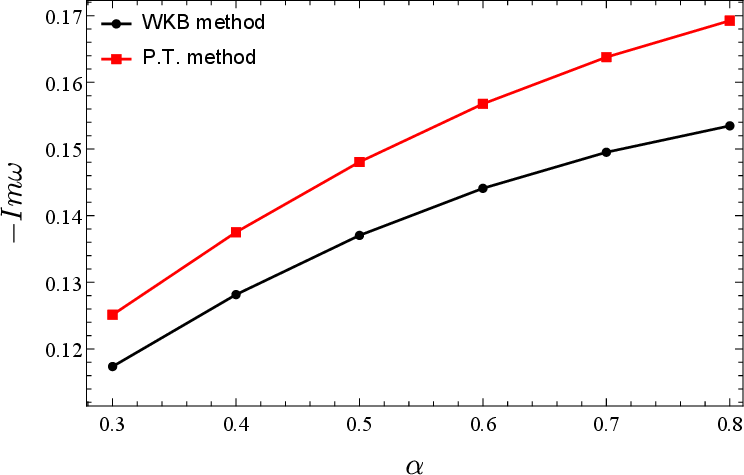}
        \label{fig:qnnb}
   \end{minipage}\ 
\caption{Qualitative comparison of the Mashhoon and WKB approximations with $M = 1$, $\sigma =0.1$, $L=1$, $n=0$,  $\omega _{q}=-0.4$ and varying values of $\alpha$ parameter. }
\label{qnmn}
\end{figure}

\newpage
\section{Conclusions} \label{sec6}
In this paper, we handled the Schwarzschild black hole surrounded by PFDM in the presence of the quintessence matter field, and we explored its thermodynamics, shadows, and quasinormal modes. To this end, we initially derived the {metric} function of the black hole and discussed the {metric} function's characteristics in four different quintessence field scenarios. We found that the quintessence matter field significantly alters the behavior of the {metric} function unless the quintessence state parameter is large enough. Then, we examined the mass function. We observed that dark matter sets a lower limit on the event horizon, but also expands its upper bound value. We noted that achieves a maximum mass value at a larger event horizon with a greater dark matter parameter. Next, we examined the black hole's thermodynamics. At first, We derived the modified Hawking temperature. We found that it tends to zero unless the quintessence state parameter is not large enough. We showed the event horizon entropy is not affected by dark matter and quintessence matter field. Then. we investigated the heat capacity function. We showed that except very large quintessence state parameter case, the black hole has a first-order phase transition. In these cases, the black hole stops exchanging radiation with its surroundings and becomes stable with a remnant mass. Then, we derived the equation of state function and discussed the pressure isotherms. Next, we investigated the shadow casts through Carter's approach, We obtained the null geodesic equations to describe the photon motion. Then, we found the effective potential and showed the impacts of dark matter and quintessence matter fields on it. We demonstrated the shadows in celestial coordinates. We found that dark matter significantly lessens the shadow radius, while quintessence matter does not. Finally, we studied the quasinormal modes with the WKB and Mashhoon approximations. We tabulated quasinormal frequencies and showed the impact of dark matter and quintessence matter fields.

\section*{Acknowledgments}
{B. C. L. is grateful to Excellence Project PřF UHK 2211/2023-2024 for the financial support.

\section*{Data Availability Statements}

The authors declare that the data supporting the findings of this study are available within the article.

\end{document}